\documentclass[useAMS,usenatbib]{mnras}
\usepackage{graphics,epsfig,longtable,lscape,txfonts,color}
\usepackage[english]{babel}

\newcommand{\mr}{\mathrm}
\newcommand{\nh}{\hbox{$N_{\mr H}$}}
\newcommand{\hcm}[1]{$\times 10^{#1}$ cm$^{-2}$}

\newcommand{\lb}{\left}
\newcommand{\rb}{\right}
\def\ie{i.\,e.}                                      
\def\eg{e.\,g.}                                      

\def\xmm{\textit{XMM-Newton}}
\def\swift{\textit{Swift}}
\def\nus{\textit{NuSTAR}}

\def\gx339{GX\,339-4}
\def\h1743{H\,1743-322}

\def\sw19{Swift\,J1910.2--0546}

\def\m31{M\,31}

\title[\nus\ view of \m31]{\nus\ view of the central region of \m31}
\author[H. Stiele, A. Kong]{H.\ Stiele$^{1}$\thanks{E-mail:
hstiele@mx.nthu.edu.tw}, A.\ K.\ H.\ Kong$^{1,2}$ \\
$^{1}$National Tsing Hua University, Department of Physics and Institute of Astronomy, No.~101 Sect.~2 Kuang-Fu Road,  30013, Hsinchu, Taiwan \\
$^{2}$Astrophysics, Department of Physics, University of Oxford, Keble Road, Oxford OX1 3RH, UK} 
\begin{document}

\date{2018 January 16}

\pagerange{\pageref{firstpage}--\pageref{lastpage}} \pubyear{2018}

\maketitle

\label{firstpage}

\begin{abstract}
Our neighbouring large spiral galaxy, the Andromeda galaxy (\m31 or NGC\,224), is an ideal target to study the X-ray source population of a nearby galaxy. \nus\ observed the central region of \m31\ in 2015 and allows studying the population of X-ray point sources at energies higher than 10 keV. Based on the source catalogue of the large \xmm\ survey of \m31, we identified counterparts to the \xmm\ sources in the \nus\ data. The \nus\ data only contain sources of a brightness comparable (or even brighter) than the selected sources that have been detected in \xmm\ data. We investigate hardness ratios, spectra and long-term light curves of individual sources obtained from \nus\ data. Based on our spectral studies we suggest four sources as possible X-ray binary candidates. The long-term light curves of seven sources that have been observed more than once show low (but significant) variability. 
\end{abstract}


\section{Introduction}
With a distance of 780 kpc \citep{1998AJ....115.1916H,1998ApJ...503L.131S} and moderate Galactic foreground absorption \citep[\nh=7\hcm{20}][]{1992ApJS...79...77S}, the Andromeda galaxy (\m31 or NGC\,224), is an ideal target to study the X-ray source population of a neighbouring large spiral galaxy. \m31 has been observed with many different X-ray imaging satellites, including \textit{Einstein} \citep{1979ApJ...234L..45V,1990ApJ...356..119C,1991ApJ...382...82T}, ROSAT \citep{1993ApJ...410..615P,1997A&A...317..328S,2001A&A...373...63S}, \textit{Chandra} \citep{2000ApJ...537L..23G,2002ApJ...570..618D,2002ApJ...578..114K,2002ApJ...577..738K,2004ApJ...610..247D,2004ApJ...609..735W,2007A&A...468...49V,2013A&A...555A..65H}, \textit{Swift} \citep{2008A&A...489..707V}, and \xmm\ \citep{2001A&A...378..800O,2010ApJ...717..739O,2005A&A...434..483P,2007A&A...465..375P,2010A&A...523A..89H,2011A&A...533A..52H,2014A&A...563A...2H}. Most of these observation focused on the central region of \m31. A large \xmm\ survey, covering the entire D$_{25}$ ellipse of \m31, has been presented by \citet{2011A&A...534A..55S}.

The X-ray source population of \m31 comprises supersoft sources \citep[the X-ray counterparts of optical novae; \eg\ ][and references therein]{2014A&A...563A...2H}, supernova remnants \citep[\eg\ ][]{2012A&A...544A.144S}, and X-ray binaries. Regarding X-ray binaries, most sources have been classified based on correlations with lists of globular clusters, fewer through the detection of long-term variability \citep[\eg\ ][]{2006ApJ...643..356W,2006ApJ...645..277T,2008A&A...480..599S,2013A&A...555A..65H}, and only a few sources through the detection of short-term variability features, characteristic for X-ray binaries \citep{2003A&A...411..553B,2005A&A...430L..45P}. In \citet{2011A&A...534A..55S} 65\% of the sources could have only been classified as ``hard" sources, \ie\ it is not possible to decide whether these sources are X-ray binaries or Crab-like supernova remnants in \m31\ or AGN in the background. This confusion is partially caused by the fact that the averaged X-ray spectra of these sources at energies below 10 keV, the ones covered by the X-ray satellites mentioned above, are quite similar.

In this paper, we make use of archival \nus\ data of \m31, that cover energies between 3 and 80 keV, and that allow us to study the X-ray emission of individual sources in \m31, to investigate the X-ray source population of \m31\ in this energy range. \citet{2016MNRAS.458.3633M} investigated joint \swift\ and \nus\ spectra of five globular cluster sources in \m31, and suggested that these sources are neutron star X-ray binaries. Our study, which comprises a bigger data sample, includes four of their sources. 

\begin{table}
\caption{Details of \nus\ observations included in this study}
\begin{center}
\begin{tabular}{lllrc}
\hline\noalign{\smallskip}
\multicolumn{1}{c}{\#} & \multicolumn{1}{c}{Obs.~id.} & \multicolumn{1}{c}{Date}  &  \multicolumn{1}{c}{Exp. [s]}  &  \multicolumn{1}{c}{Field}\\
 \hline\noalign{\smallskip}
1 & 50026001002	& 2015-02-06 	& 106386	&M1\\
2 & 50026001004	& 2015-03-01 	& 104295	&M1\\
3 & 50111001002	& 2015-06-26 	& 104123	&A\\
4 & 50110001002	& 2015-07-25 	& 52232	&A\\
5 & 50110001004	& 2015-08-01 	& 71376	&A\\
6 & 50101001002	& 2015-09-13 	& 98550	&Bulge\\
\hline\noalign{\smallskip}  
\end{tabular} 
\end{center}
\label{Tab:data}
\end{table}

\begin{table*}
\caption{List of sources included in this study}
\begin{center}
\begin{tabular}{rlllrrcl}
\hline\noalign{\smallskip}
 \multicolumn{1}{c}{Scr.\ ID} & \multicolumn{1}{c}{R.A.} & \multicolumn{1}{c}{Dec.}  & \multicolumn{1}{c}{P$_{\mr{err}}$} &  \multicolumn{1}{c}{L$^{\dagger}$} & \multicolumn{1}{c}{L$_{\mr{err}}^{\dagger}$} & \multicolumn{1}{c}{Class}  & \multicolumn{1}{c}{Remark}\\
 \multicolumn{1}{c}{[SPH11]} & &  & \multicolumn{1}{c}{arcsec} &  \multicolumn{1}{c}{erg/s} & \multicolumn{1}{c}{erg/s} &   &\\
 \hline\noalign{\smallskip}
807 &	00	42	9.53 & +41 17 44.6 & 1.81 & 3.54E+35 & 1.16E+34 & $<$GlC$>$	&mita 140	 \\
\emph{823}	& 00 42 12.17	& +41 17 57.5	& 1.78	& 6.60E+35	& 1.16E+34	& $<$GlC$>$   	& Bo 078       \\                                                                                                                                                                                                                          
829 &	00	42	13.17	&+41	18	35.1	&1.73&	3.60E+36	&2.90E+34&	$<$hard$>$	&[SK98] 194 \\
836 &	00	42	15.28	&+41	18	0.9	&2.24	&5.85E+34	&4.64E+33	&$<$hard$>$	 &\\
\emph{855}	& 00 42 18.71	& +41 14 01.0	& 1.73	& 2.44E+36	& 2.32E+34	& GlC      	& Bo 086          \\                                                                                                                                                                                                                                                       
\emph{870}	& 00 42 21.51	& +41 16 00.3	& 1.74	& 1.33E+36	& 1.74E+34	& $<$hard$>$   	&   	\\
872 &	00	42	21.68& +41	14	18.2&	1.82	&2.90E+35	&5.79E+33&	$<$GlC$>$	& type I X-ray burst [PH2005] \\
\emph{876}	& 00 42 22.51	& +41 13 33.5	& 1.74	& 9.85E+35	& 1.16E+34	& $<$hard$>$   	&  	\\
\emph{877}	& 00 42 23.00	& +41 15 34.3	& 1.72	& 6.20E+36	& 2.90E+34	& $<$hard$>$   	& XRB in [PFH2005]    \\                                                                                                                                                                                                                                                                    
879 &00	42	23.2&	+41	14	6.5&	1.79	&2.06E+35&	5.79E+33&	$<$hard$>$& \\
\emph{910}	& 00 42 28.31	& +41 09 59.9	& 1.75	& 1.20E+36	& 1.74E+34	& $<$XRB$>$    	&                                    \\                                                                                                                                                                                                               
\emph{911}	 	& 00 42 28.37	& +41 12 22.6	& 1.72	& 3.18E+36	& 2.32E+34	& $<$hard$>$   	& \\
\emph{922}	& 00 42 31.16	& +41 16 20.7	& 1.73	& 2.90E+36	& 2.32E+34	& $<$hard$>$   	& \\
\emph{930}	& 00 42 32.21	& +41 13 13.5	& 1.72	& 1.82E+36	& 1.74E+34	& $<$hard$>$   	& \\
940  	&	00   42   33.93  & +41   16   18.8 &  1.8  & 3.51E+35  & 5.79E+33  & $<$hard$>$   &     \\
\emph{966}	& 00 42 38.61	& +41 16 03.0	& 1.72	& 2.03E+37	& 5.21E+34	& XRB       	& Z-source LMXRB [BKO2003]   \\                                                                                                                                                                                                                                                    \emph{972}	& 00 42 40.22	& +41 18 44.9	& 1.8	& 2.86E+35	& 5.79E+33	& $<$hard$>$   	&\\ 	
981  	&	00   42   41.49 &  +41   15   24.2 &  1.88  & 2.92E+35 &  1.16E+34 &  GlC  & Bo 124   \\
985  	&	00   42   41.85 &  +41   16   35.8  & 1.74  & 4.03E+36  & 4.06E+34 &  XRB &  LMXRB, X-ray transient, BH candidate [WGC2006]   [TPC2006] \\
990  	&	00   42   42.18  & +41   16   7.5 &  1.76  & 2.94E+36  & 5.79E+34 &  XRB &  X-ray transient, BH candidate [OBT2001] [TBP2001]    \\
994  	&	00   42   42.5 &  +41   15   51.7  & 1.75 &  8.63E+35 &  2.32E+34  & $<$hard$>$ &  \\
997  	&	00   42   42.92  & +41   15   41.7 &  1.94  & 2.45E+35 &  1.16E+34 &  $<$hard$>$ &   \\
1005	&	00   42   43.82  & +41   16   31.2 &  1.84 &  5.04E+35  & 1.16E+34 &  $<$XRB$>$  & recurrent X-ray transient \\
1010	&	00   42   44.48  &+ 41   16   8.4  & 1.79 &  6.20E+35 &  2.32E+34  & $<$XRB$>$  & recurrent X-ray transient; nucleus  \\
\emph{1015}	&	00   42   45  & +41   16   19.6  & 1.85  & 7.30E+35  & 1.74E+34 &  $<$hard$>$  & WSTB 37W 135   \\
\emph{1024}	& 00 42 47.16	& +41 16 27.2	& 1.72	& 4.83E+36	& 2.90E+34	& $<$XRB$>$    	&   \\    
\emph{1036}	& 00 42 48.50	& +41 15 22.2	& 1.73	& 4.72E+36	& 2.90E+34	& $<$hard$>$   	& \\
1041	&	00   42   49.24 &  +41   18   16  & 1.84  & 2.16E+35 &  5.79E+33  & $<$hard$>$  &  \\
1059	&	00   42   52.44  & +41   16   49.1  & 1.94  & 3.49E+35 &  1.74E+34 &  $<$XRB$>$  &   X-ray transient    \\
\emph{1060}	& 00 42 52.50	& +41 18 54.3	& 1.73	& 3.30E+36	& 2.32E+34	& $<$XRB$>$    	& superburst, NS LMXRB [K2002]  \\                                                                                                                                                                                                                       \emph{1075}	& 00 42 54.98	& +41 16 03.5	& 1.73	& 3.48E+36	& 2.32E+34	& $<$hard$>$   	&  \\ 	
1078	&	00   42   55.39 &  +41   18   35.7  & 1.79  & 3.34E+35 &  5.79E+33 &  $<$hard$>$ &\\
1084	&	00   42   56.89 &  +41   18   43.2  & 2.97  & 4.62E+35 &  4.64E+34 &  $<$XRB$>$ &  X-ray transient \\
1091	&	00   42   58.01 &  +41   11     5.0  & 1.74  & 1.66E+36 &  1.74E+34 &  $<$hard$>$ &   \\
1093	&	00   42   58.36 &  +41   15   29.6  & 1.83  & 2.57E+35 &  5.79E+33 &  $<$hard$>$ & \\
\emph{1102}	& 00 42 59.65	& +41 19 19.4	& 1.73	& 1.87E+36	& 1.74E+34	& GlC      	& Bo 143  \\                                                                                                                    
\emph{1103}	& 00 42 59.90	& +41 16 05.9	& 1.73	& 1.92E+36	& 1.74E+34	& GlC      	& Bo 144 \\                                                                                                                     
\emph{1116}	& 00 43 03.08	& +41 15 25.0	& 1.73	& 2.85E+36	& 2.32E+34	& GlC      	& Bo 146  \\                                                                                                                    
\emph{1122}	& 00 43 03.89	& +41 18 05.4	& 1.73	& 1.44E+36	& 1.74E+34	& GlC      	& Bo 148   \\                                                                                                                   
\emph{1124} 	&	00   43   4.27 &  +41   16   1.6  & 1.86 &  2.13E+35  & 5.79E+33 &  $<$GlC$>$ & \\
1136	&	00   43   7.1  & +41   18   10.4  & 1.81 &  4.87E+35  & 1.74E+34 &  $<$XRB$>$  &\\
\emph{1157}	& 00 43 10.66	& +41 14 51.8	& 1.72	& 4.40E+36	& 2.90E+34	& GlC      	& Bo 153  \\                                                                                                                                                                                                                                         
\emph{1253}	& 00 43 32.44	& +41 10 41.3	& 1.75	& 1.45E+36	& 2.32E+34	& $<$hard$>$   	& \\
1257  	&	00   43   33.9 & + 41   14   6.5  & 4.51 &  3.70E+34 &  5.79E+33   & $<$hard$>$   	&   \\
\emph{1267}	& 00 43 37.30	& +41 14 43.9	& 1.74	& 2.78E+36	& 2.90E+34	& GlC      	& Bo 185  \\                                                                                                              
\hline\noalign{\smallskip} 
\end{tabular} 
\end{center}
Notes:\\
values and classification taken from \citet{2011A&A...534A..55S}\\
WGC2006: \citet{2006ApJ...637..479W}, TPC2006: \citet{2006ApJ...645..277T}, PFH2005: \citet{2005A&A...434..483P}, PH2005: \citet{2005A&A...430L..45P}, BKO2003: \citet{2003A&A...411..553B}, K2002: \citet{2002ApJ...578..114K}, OBT2001: \citet{2001A&A...378..800O}, TBP2001: \citet{2001ApJ...563L.119T}, SK98: \citet{1998AstL...24...69S}, mita: \citet{1993PhDT........41M}, WSTB 37W: \citet{1985A&AS...61..451W}\\
$^{\dagger}$: luminosity in the 0.2 -- 4.5 keV band, assuming a distance to \m31\ of 780 kpc\\
italic source IDs indicate sources for which we found \nus\ counterparts
\label{Tab:srcs}
\end{table*}

\begin{table*}
\caption{Count rate ratios and significance of sources included in this study}
\begin{center}
\begin{tabular}{rrrrr}
\hline\noalign{\smallskip}
 \multicolumn{1}{c}{Scr.\ ID} & \multicolumn{2}{c}{region 15\arcsec} & \multicolumn{2}{c}{region 30\arcsec}\\
 \multicolumn{1}{c}{[SPH11]} & \multicolumn{1}{c}{$R$} &  \multicolumn{1}{c}{Sig} & \multicolumn{1}{c}{$R$} & \multicolumn{1}{c}{Sig}\\
 \hline\noalign{\smallskip}
807   &$ 1.195 \pm  0.001  $&$  1.824 \pm  0.089  $&$ 1.255 \pm  0.043 $&$   4.277 \pm  0.806  $\\ 
823   &$ 1.313 \pm  0.006  $&$  2.942 \pm  0.094  $&$ 1.326 \pm  0.018 $&$   5.440 \pm  0.164  $\\ 
829   &$ 1.152 \pm  0.018  $&$  1.418 \pm  0.166  $&$ 1.186 \pm  0.067 $&$   3.144 \pm  0.984  $\\ 
836   &$ 0.984 \pm  0.103  $&$ -0.209 \pm  0.916  $&$ 0.986 \pm  0.024 $&$  -0.271 \pm  0.446  $\\ 
855   &$ 1.864 \pm  0.025  $&$  8.034 \pm  0.268  $&$ 1.620 \pm  0.039 $&$  10.637 \pm  0.516  $\\ 
870   &$ 1.313 \pm  0.102  $&$  3.371 \pm  1.015  $&$ 1.676 \pm  0.213 $&$  10.643 \pm  2.415  $\\ 
872   &$ 1.040 \pm  0.086  $&$  0.400 \pm  0.909  $&$ 1.054 \pm  0.074 $&$   1.066 \pm  1.494  $\\ 
876   &$ 1.639 \pm  0.112  $&$  5.837 \pm  0.710  $&$ 1.223 \pm  0.001 $&$   4.480 \pm  0.103  $\\ 
877   &$ 1.247 \pm  0.080  $&$  2.586 \pm  0.807  $&$ 1.532 \pm  0.128 $&$   8.810 \pm  1.864  $\\ 
879   &$ 0.895 \pm  0.084  $&$ -1.281 \pm  1.049  $&$ 1.042 \pm  0.009 $&$   0.910 \pm  0.200  $\\ 
910   &$ 1.836 \pm  0.401  $&$  6.196 \pm  2.279  $&$ 1.562 \pm  0.077 $&$   8.134 \pm  0.835  $\\ 
911   &$ 1.418 \pm  0.115  $&$  4.042 \pm  0.984  $&$ 1.132 \pm  0.020 $&$   2.711 \pm  0.320  $\\ 
922   &$ 1.444 \pm  0.151  $&$  5.477 \pm  1.693  $&$ 1.669 \pm  0.237 $&$  12.429 \pm  3.343  $\\ 
930   &$ 2.112 \pm  0.023  $&$ 15.820 \pm  0.387  $&$ 3.211 \pm  0.319 $&$  33.918 \pm  3.003  $\\ 
940   &$ 1.025 \pm  0.010  $&$  0.358 \pm  0.128  $&$ 0.791 \pm  0.049 $&$  -7.417 \pm  1.895  $\\ 
966   &$ 2.347 \pm  0.101  $&$ 22.321 \pm  1.152  $&$ 3.116 \pm  0.252 $&$  41.902 \pm  2.000  $\\       
971   &$ 0.826 \pm  0.026  $&$ -3.955 \pm  0.467  $&$ 1.748 \pm  0.054 $&$  20.122 \pm  1.663  $\\      
972   &$ 1.341 \pm  0.203  $&$  2.677 \pm  1.484  $&$ 1.215 \pm  0.016 $&$   3.385 \pm  0.213  $\\ 
981   &$ 0.856 \pm  0.042  $&$ -2.025 \pm  0.454  $&$ 0.686 \pm  0.077 $&$ -11.287 \pm  2.774  $\\ 
985   &$ 0.848 \pm  0.053  $&$ -2.852 \pm  1.126  $&$ 0.841 \pm  0.053 $&$  -6.326 \pm  2.197  $\\ 
990   &$ 1.057 \pm  0.021  $&$  1.063 \pm  0.356  $&$ 1.073 \pm  0.002 $&$   2.664 \pm  0.101  $\\ 
994   &$ 1.015 \pm  0.032  $&$  0.276 \pm  0.568  $&$ 0.907 \pm  0.057 $&$  -3.593 \pm  2.226  $\\ 
997   &$ 0.997 \pm  0.017  $&$ -0.029 \pm  0.267  $&$ 0.858 \pm  0.055 $&$  -5.136 \pm  1.961  $\\ 
1005  &$ 1.085 \pm  0.039  $&$  1.465 \pm  0.669  $&$ 1.184 \pm  0.022 $&$   5.810 \pm  0.704  $\\ 
1010  &$ 1.029 \pm  0.050  $&$  0.527 \pm  0.910  $&$ 1.264 \pm  0.053 $&$   8.029 \pm  1.468  $\\ 
1015  &$ 1.178 \pm  0.002  $&$  3.071 \pm  0.020  $&$ 1.427 \pm  0.020 $&$  12.089 \pm  0.474  $\\ 
1024  &$ 1.418 \pm  0.059  $&$  6.279 \pm  0.905  $&$ 1.526 \pm  0.111 $&$  13.179 \pm  2.331  $\\ 
1036  &$ 1.533 \pm  0.182  $&$  4.390 \pm  1.602  $&$ 1.398 \pm  0.328 $&$   6.568 \pm  4.882  $\\ 
1041  &$ 1.007 \pm  0.215  $&$  0.136 \pm  1.337  $&$ 0.948 \pm  0.162 $&$  -0.755 \pm  2.408  $\\ 
1059  &$ 1.062 \pm  0.124  $&$  0.373 \pm  0.981  $&$ 0.908 \pm  0.064 $&$  -1.833 \pm  1.407  $\\ 
1060  &$ 1.529 \pm  0.175  $&$  4.387 \pm  1.445  $&$ 1.613 \pm  0.154 $&$   8.385 \pm  2.127  $\\ 
1075  &$ 1.404 \pm  0.124  $&$  3.443 \pm  1.044  $&$ 1.201 \pm  0.054 $&$   3.455 \pm  1.121  $\\ 
1078  &$ 1.047 \pm  0.101  $&$  0.407 \pm  0.874  $&$ 1.061 \pm  0.152 $&$   0.720 \pm  2.606  $\\ 
1084  &$ 0.991 \pm  0.130  $&$ -0.171 \pm  1.103  $&$ 0.937 \pm  0.150 $&$  -1.433 \pm  2.594  $\\ 
1091  &$ 1.251 \pm  0.183  $&$  2.016 \pm  1.387  $&$ 1.342 \pm  0.031 $&$   5.118 \pm  0.404  $\\ 
1093  &$ 0.998 \pm  0.115  $&$ -0.077 \pm  1.088  $&$ 0.936 \pm  0.131 $&$  -1.432 \pm  2.935  $\\ 
1102  &$ 1.439 \pm  0.174  $&$  3.586 \pm  1.274  $&$ 1.391 \pm  0.202 $&$   5.719 \pm  2.399  $\\ 
1103  &$ 1.500 \pm  0.200  $&$  4.246 \pm  1.329  $&$ 1.309 \pm  0.065 $&$   5.242 \pm  1.016  $\\ 
1116  &$ 1.405 \pm  0.128  $&$  3.557 \pm  1.116  $&$ 1.323 \pm  0.222 $&$   5.315 \pm  3.527  $\\ 
1122  &$ 1.468 \pm  0.167  $&$  3.943 \pm  1.408  $&$ 1.476 \pm  0.121 $&$   7.069 \pm  2.107  $\\ 
1124  &$ 1.529 \pm  0.175  $&$  4.387 \pm  1.445  $&$ 1.613 \pm  0.154 $&$   8.385 \pm  2.127  $\\ 
1136  &$ 1.016 \pm  0.123  $&$  0.011 \pm  1.017  $&$ 1.021 \pm  0.084 $&$   0.207 \pm  1.453  $\\ 
1157  &$ 2.216 \pm  0.246  $&$ 11.073 \pm  1.440  $&$ 2.402 \pm  0.213 $&$  18.653 \pm  1.955  $\\ 
1253  &$ 1.878 \pm  0.233  $&$  6.005 \pm  1.675  $&$ 2.127 \pm  0.327 $&$  11.336 \pm  3.689  $\\ 
1257  &$ 0.971 \pm  0.086  $&$ -0.320 \pm  0.741  $&$ 0.891 \pm  0.067 $&$  -1.885 \pm  1.163  $\\ 
1267  &$ 1.848 \pm  0.197  $&$  7.256 \pm  1.293  $&$ 2.017 \pm  0.224 $&$  13.014 \pm  1.751  $\\ 
\hline\noalign{\smallskip} 
\end{tabular} 
\end{center}
\label{Tab:stats}
\end{table*}

\begin{table*}
\caption{List of background subtracted \nus\ FPMA/B source counts (in units of $\times 10^{-3}$ cts/s) from a source extraction region of 30\arcsec.}
\begin{center}
\small
\begin{tabular}{rccccccc}
\hline\noalign{\smallskip}
 \multicolumn{1}{c}{[SPH11]} & \multicolumn{1}{c}{FPM}  & \multicolumn{1}{c}{obs.\ 1} &  \multicolumn{1}{c}{obs.\ 2} & \multicolumn{1}{c}{obs.\ 3} & \multicolumn{1}{c}{obs.\ 4} & \multicolumn{1}{c}{obs.\ 5} & \multicolumn{1}{c}{obs.\ 6}\\
 \hline\noalign{\smallskip}
823 &	A &		--		 &	--		   &	--		     &	--		        &	--		    & $3.28\pm 0.32$\\  
    &	B &		--		 &	--		   &	--		     &	--		        &	--		    & $3.01\pm 0.31$\\
855 &		A &	--		 &	--		   &	--		     &	--		        &	--		    & $5.83\pm 0.39$\\  
    &		B &	--		 &	--		   &	--		     &	--		        &	--		    & $5.44\pm 0.39$\\
870 &		A &	--		 &	--		   &	--		     &	--		        &	--		    & $5.44\pm 0.36$\\  
    &		B &	--		 &	--		   &	--		     &	--		        &	--		    & $5.93\pm 0.35$\\
876 &		A &	--		 &	--		   &	--		     &	--		        &	--		    & $4.49\pm 0.37$\\  
    &		B &	--		 &	--		   &	--		     &	--		        &	--		    & $4.03\pm 0.36$\\
877 &		A &	--		 &	--		   &	--		     &	--		        &	--		    & $4.43\pm 0.39$\\  
    &		B &	--		 &	--		   &	--		     &	--		        &	--		    & $3.66\pm 0.35$\\
910 &		A &	--		 &	--		   &	--		     &	--		        &	--		    & $2.55\pm 0.33$\\ 
    &		B &	--		 &	--		   &	--		     &	--		        &	--		    & $2.75\pm 0.32$\\
911 &	A &		--		 &	--		   &	--		     &	--		        &	--		    & $4.10\pm 0.37$\\  
    &	B &		--		 &	--		   &	--		     &	--		        &	--		    & $3.12\pm 0.36$\\
922 &	A &		--		 &	--		   &	--		     &	--		        &	--		    & $8.49\pm 0.42$\\ 
    &	B &		--		 &	--		   &	--		     &	--		        &	--		    & $9.44\pm 0.45$\\
930 &	A &		--		 &	--		   &	--		     &	--		        &	--		    & $23.52\pm 0.57$\\ 
    &	B &		--		 &	--		   &	--		     &	--		        &	--		    & $27.23\pm 0.61$\\
966 &	A &		--		 &	--		   &	--		     &	--		        &	--		    & $40.03\pm 0.71$\\ 
    &	B &		--		 &	--		   &	--		     &	--		        &	--		    & $39.24\pm 0.71$\\
972 &	A &		--		 &	--		   &	--		     &	--		        &	--		    & $1.08\pm 0.29$\\ 
    &	B &		--		 &	--		   &	--		     &	--		        &	--		    & $1.20\pm 0.29$\\
1015& 	A &		--		 &	--		   &	--		     & $5.73\pm 0.54$ &  $5.30\pm 0.45$ & $19.82\pm 0.53$\\ 
    & 	B &		--		 &	--		   &	--		     & $2.83\pm 0.37$ &  $2.74\pm 0.31$ & $19.83\pm 0.53$\\
1024& 	A &		--		 &	--		   &	--		     & $6.32\pm 0.55$ &  $5.40\pm 0.47$ & $15.92\pm 0.49$\\ 
    & 	B &		--		 &	--		   &	--		     & $4.62\pm 0.49$ &  $4.29\pm 0.42$ & $17.71\pm 0.51$\\
1036&	A &	$5.14\pm 0.34$ &	--		   &	--		     & $3.90\pm 0.45$ &  $3.88\pm 0.38$ & $10.99\pm 0.44$\\ 
    &   B &	$7.31\pm 0.37$ &	--		   &	--		     & $1.09\pm 0.24$ &  $1.14\pm 0.25$ & $9.12\pm 0.40$\\
1060&	A &		--		 &	--		   &$2.72\pm 0.38$ & $4.03\pm 0.52$ &  $3.60\pm 0.43$ & $4.91\pm 0.35$\\ 
    &	B &		--		 &	--	           &$3.32\pm 0.37$ & $2.89\pm 0.41$ &  $2.30\pm 0.30$ & $4.48\pm 0.36$\\
1075&	A &	$2.01\pm 0.30$ &	--		   &	--		     & $2.65\pm 0.50$ &  $1.45\pm 0.41$ & $4.36\pm 0.36$\\ 
    &   B &	$3.08\pm 0.31$ &        --                &         --               & $3.38\pm 0.48$ &  $2.40\pm 0.42$ & $4.47\pm 0.36$\\
1091&	A &		--		 &	--		   &	--		     &	--		        &         --                 & $1.69\pm 0.29$\\ 
    &     B &	       --               &         --              &          --              &       --                  &         --                 & $1.86\pm 0.30$\\
1102&	A &	$1.45\pm 0.28$ &$1.45\pm 0.31$ & $2.51\pm 0.38$&  $1.57\pm 0.47$&  $1.41\pm 0.41$ & $1.65\pm 0.30$\\ 
    &   B &	$2.21\pm 0.31$ &$2.10\pm 0.34$ & $2.13\pm 0.36$&  $1.45\pm 0.46$&  $3.00\pm 0.40$ & $2.40\pm 0.32$\\
1103&	A &	$3.20\pm 0.29$ &$3.38\pm 0.32$ & $4.07\pm 0.36$&  $5.09\pm 0.52$&  $5.11\pm 0.44$ & $4.77\pm 0.36$\\ 
    &   B &	$3.83\pm 0.31$ &$3.91\pm 0.34$ & $2.29\pm 0.27$&  $5.00\pm 0.49$&  $4.75\pm 0.41$ & $6.01\pm 0.38$\\
1116&	A &	$2.25\pm 0.28$ &$4.09\pm 0.33$ & $3.05\pm 0.34$&  $3.67\pm 0.49$&  $4.63\pm 0.43$ & $4.90\pm 0.36$\\ 
    &   B &	$2.19\pm 0.28$ &$6.02\pm 0.36$ & $1.53\pm 0.26$&  $3.21\pm 0.45$&  $3.59\pm 0.39$ & $5.63\pm 0.36$\\
1122&	A &	$2.17\pm 0.30$ &$1.53\pm 0.31$ & $4.29\pm 0.40$&  $1.90\pm 0.48$&  $2.10\pm 0.42$ & $2.31\pm 0.31$\\ 
    &   B &	$3.27\pm 0.33$ &$1.82\pm 0.32$ & $4.01\pm 0.38$&  $2.18\pm 0.48$&  $2.31\pm 0.39$ & $2.32\pm 0.32$\\
1124&	A &	$1.24\pm 0.26$ &$1.96\pm 0.30$ & $1.72\pm 0.34$&  $1.93\pm 0.45$&  $3.64\pm 0.37$ & $2.19\pm 0.32$\\ 
    &   B &	$1.30\pm 0.26$ &$1.72\pm 0.30$ & $1.54\pm 0.33$&  $2.13\pm 0.43$&  $3.28\pm 0.38$ & $4.62\pm 0.36$\\
1157&	A &	$7.82\pm 0.36$ &$8.26\pm 0.39$ & $13.63\pm 0.47$&  $13.98\pm 0.65$&  $12.74\pm 0.54$ &--\\
    &   B &	$8.46\pm 0.36$ &$9.18\pm 0.39$ & $11.64\pm 0.44$&  $12.70\pm 0.63$&  $12.78\pm 0.54$ &--\\
1253&	A &	$4.63\pm 0.28$ &	--	           & $2.26\pm 0.23$&  $5.10\pm 0.40$&  $4.46\pm 0.33$ &--\\
    &   B &	$5.25\pm 0.31$ &        --                & $6.09\pm 0.37$&  $2.52\pm 0.26$&  $2.73\pm 0.25$ &--\\
1267&	A &	$5.38\pm 0.34$ &$5.56\pm 0.35$ & $6.85\pm 0.37$&  $5.96\pm 0.47$&  $7.04\pm 0.44$ &--\\ 
    &   B &	$6.29\pm 0.32$ &$5.51\pm 0.34$ & $7.25\pm 0.39$&  $6.96\pm 0.52$&  $6.58\pm 0.45$ &--\\
\hline\noalign{\smallskip} 
\end{tabular} 
\end{center}
\label{Tab:nuscts}
\end{table*}
   
\begin{figure*}
\resizebox{0.35\hsize}{!}{\includegraphics[clip,angle=0]{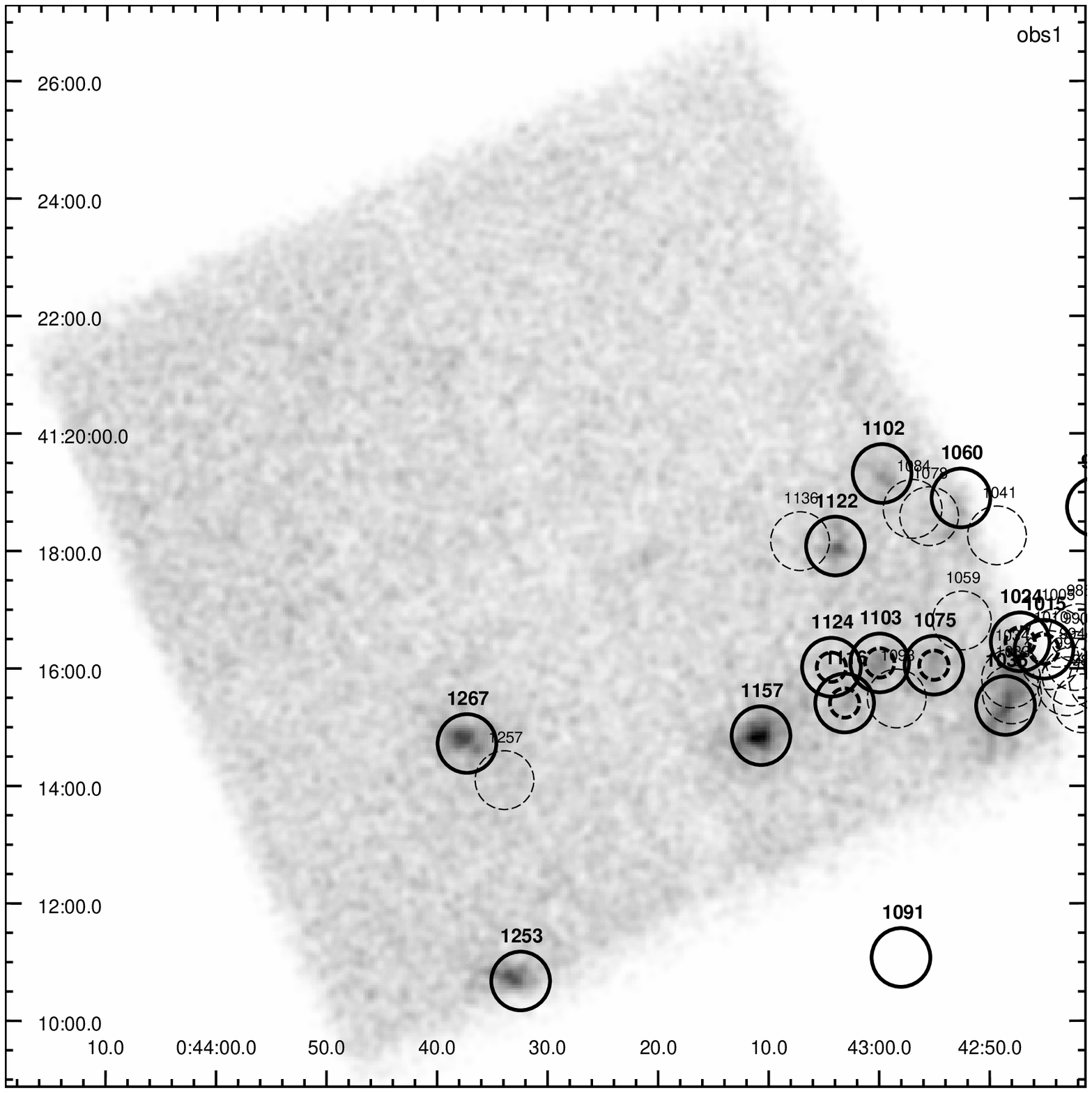}}\vspace{0.5cm}\resizebox{0.35\hsize}{!}{\includegraphics[clip,angle=0]{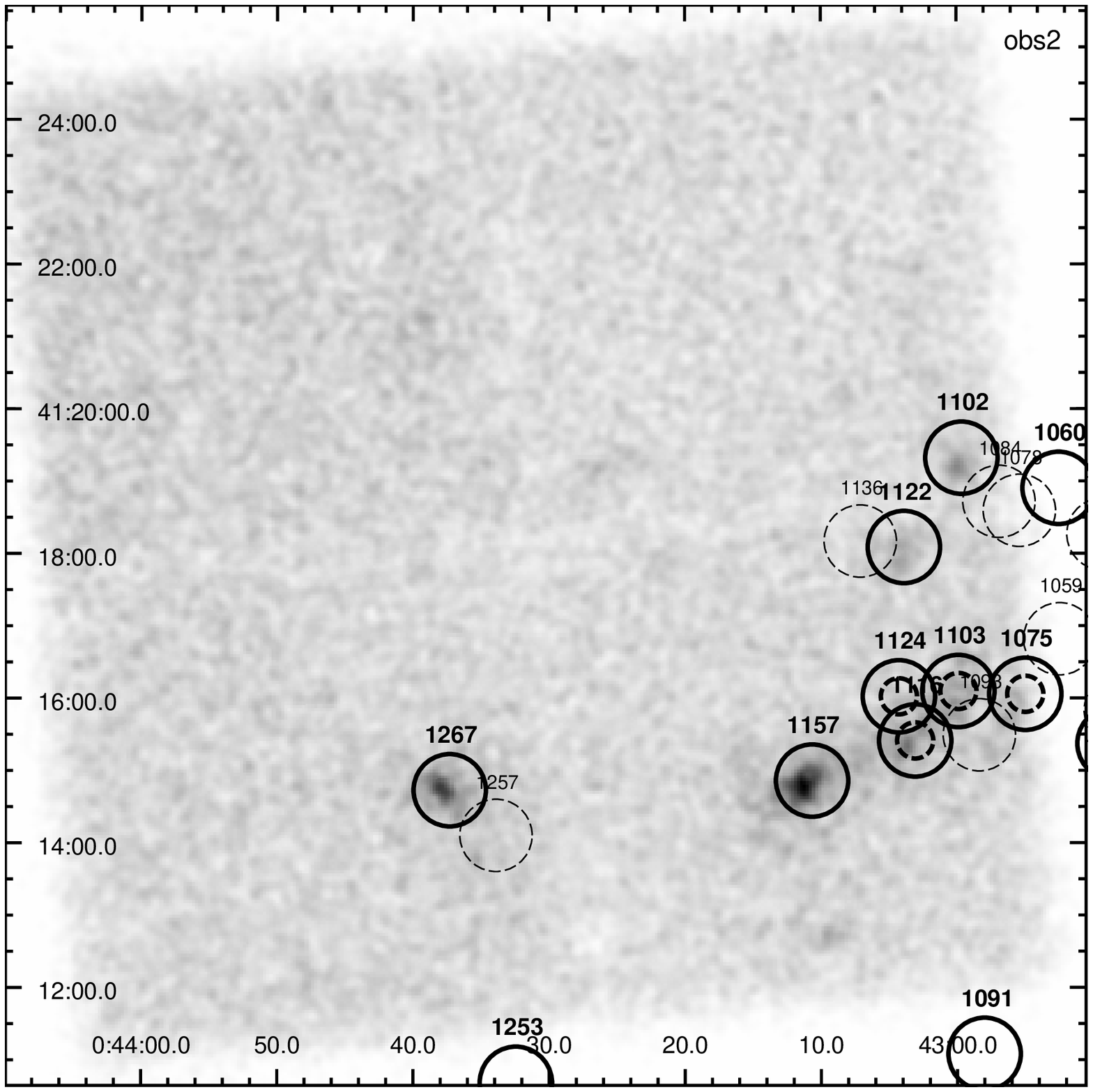}}\\
\resizebox{0.35\hsize}{!}{\includegraphics[clip,angle=0]{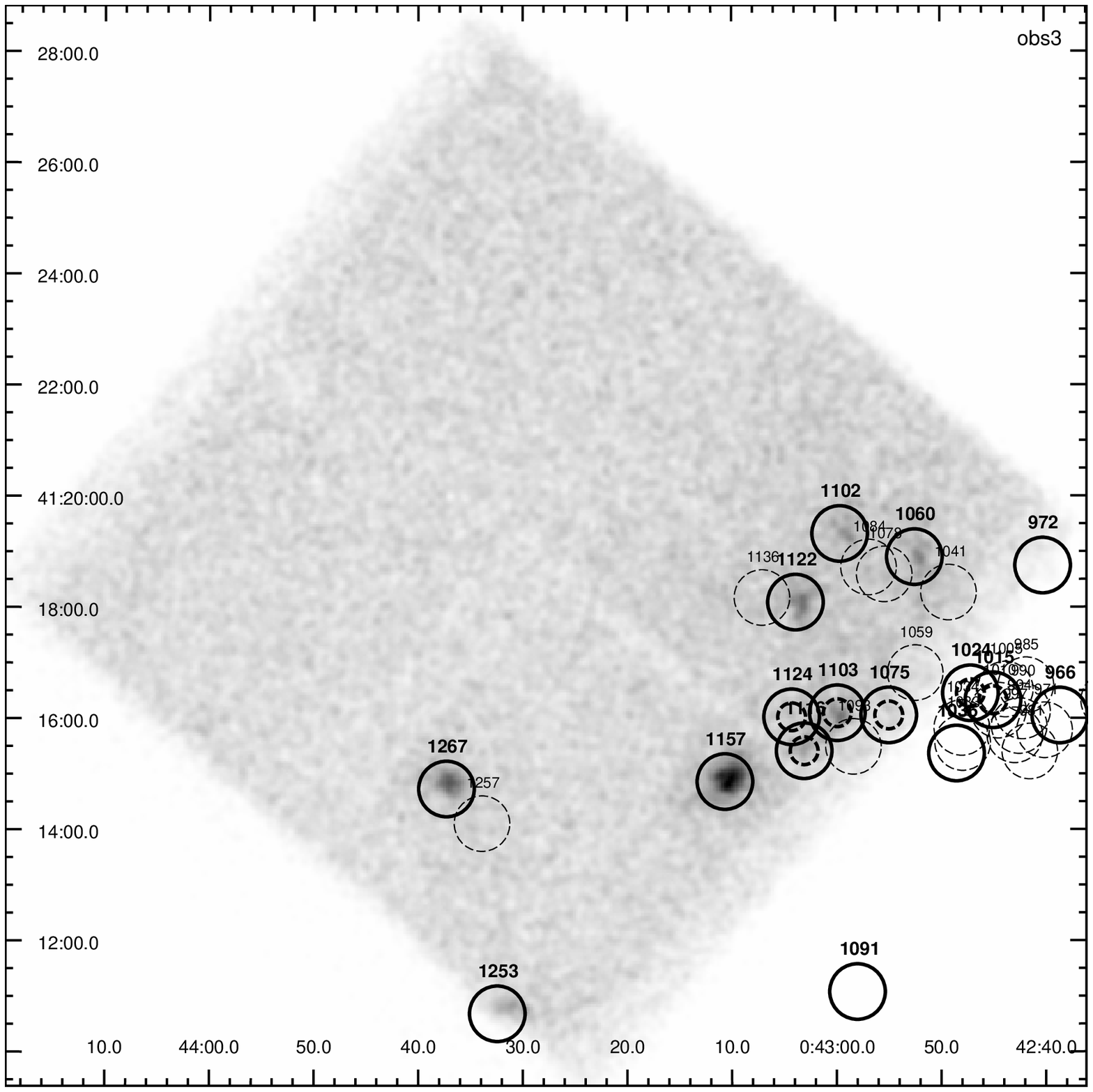}}\vspace{0.5cm}\resizebox{0.35\hsize}{!}{\includegraphics[clip,angle=0]{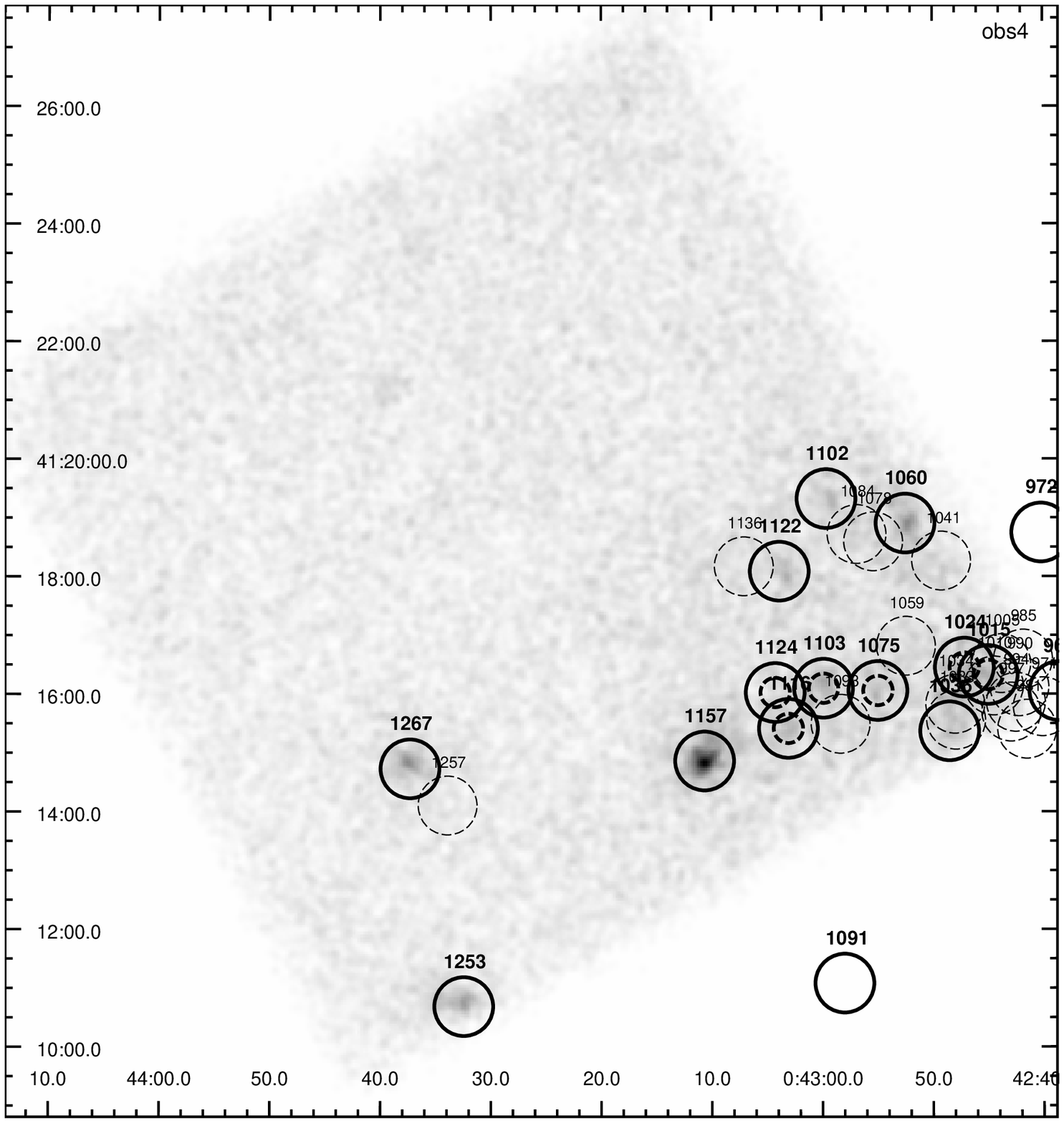}}\\
\resizebox{0.35\hsize}{!}{\includegraphics[clip,angle=0]{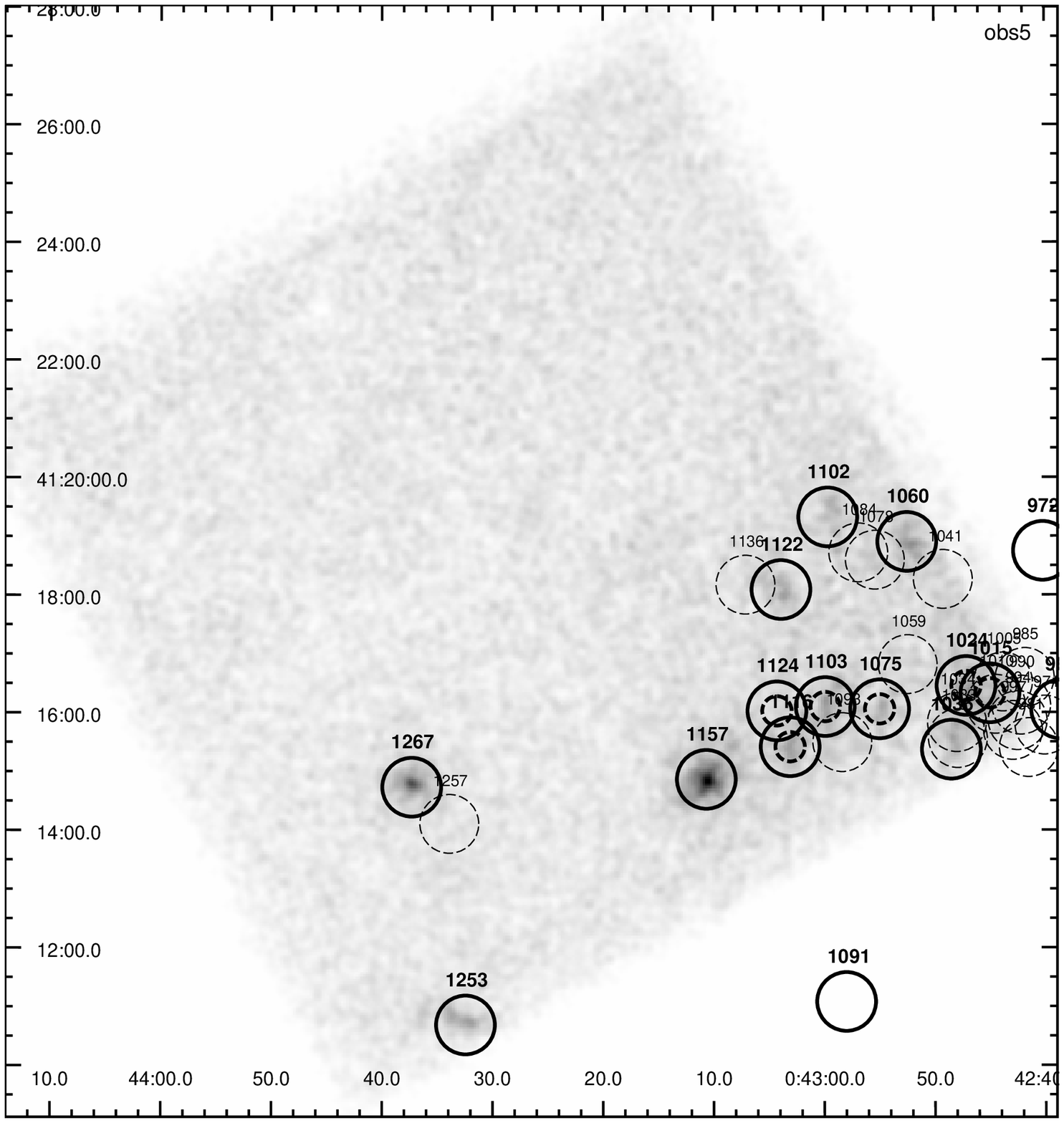}}\vspace{0.5cm}\resizebox{0.35\hsize}{!}{\includegraphics[clip,angle=0]{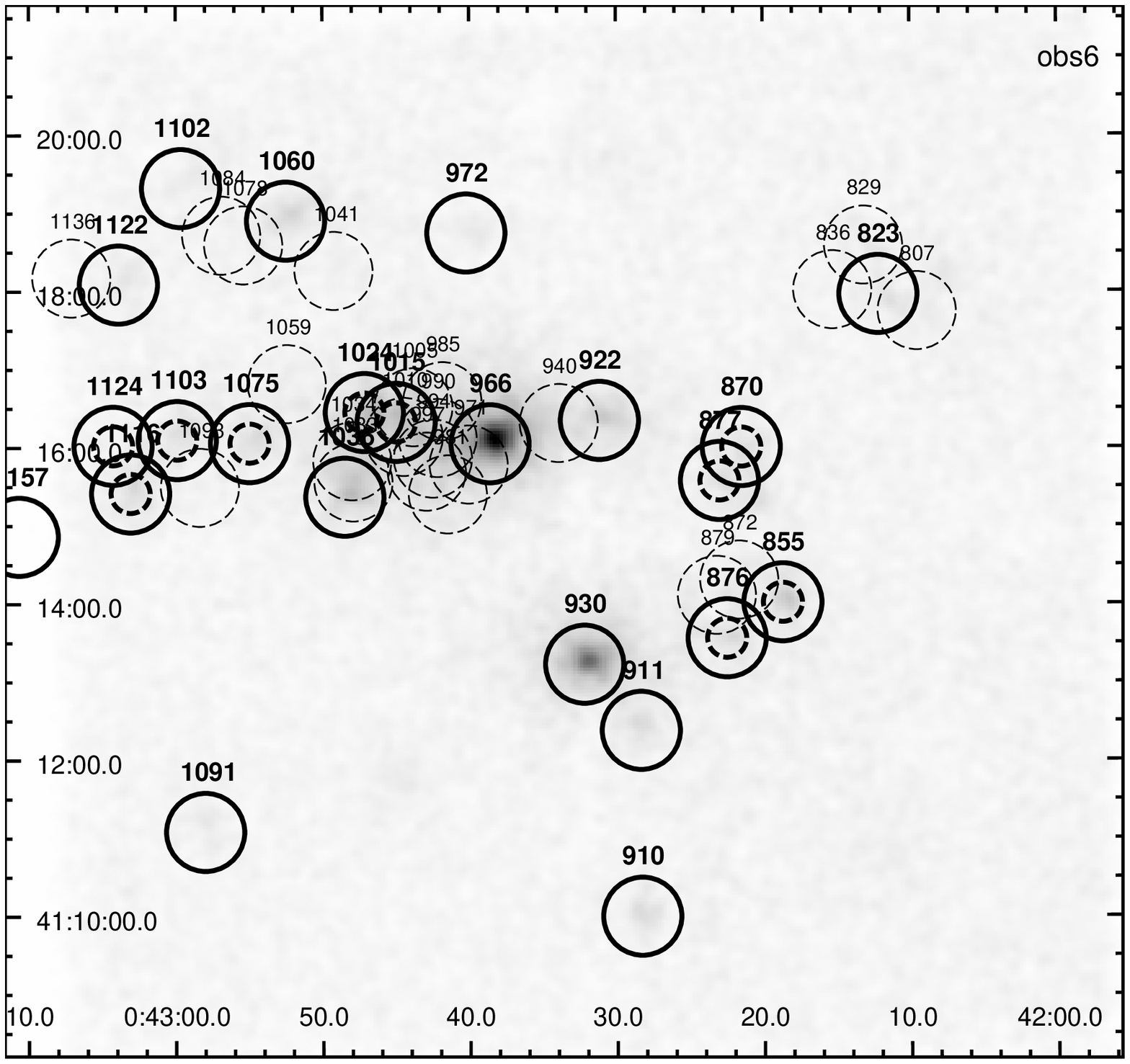}}\\
\caption{Images of the \nus/FPMA data. Sources detected by \xmm\ \citep{2011A&A...534A..55S} for which we found a \nus\ counterpart (see Sect.~\ref{Sec:obs}) are indicated by black circles. Nearby \xmm\ sources, which do not have a \nus\ counterpart according to our selection criteria are marked as dashed circles. For sources with \nus\ counterpart where source regions of 30\arcsec overlap each other, we also indicate regions of 15\arcsec using dashed circles.}
\label{Fig:ima_nusa}
\end{figure*}

\begin{figure*}
\resizebox{0.35\hsize}{!}{\includegraphics[clip,angle=0]{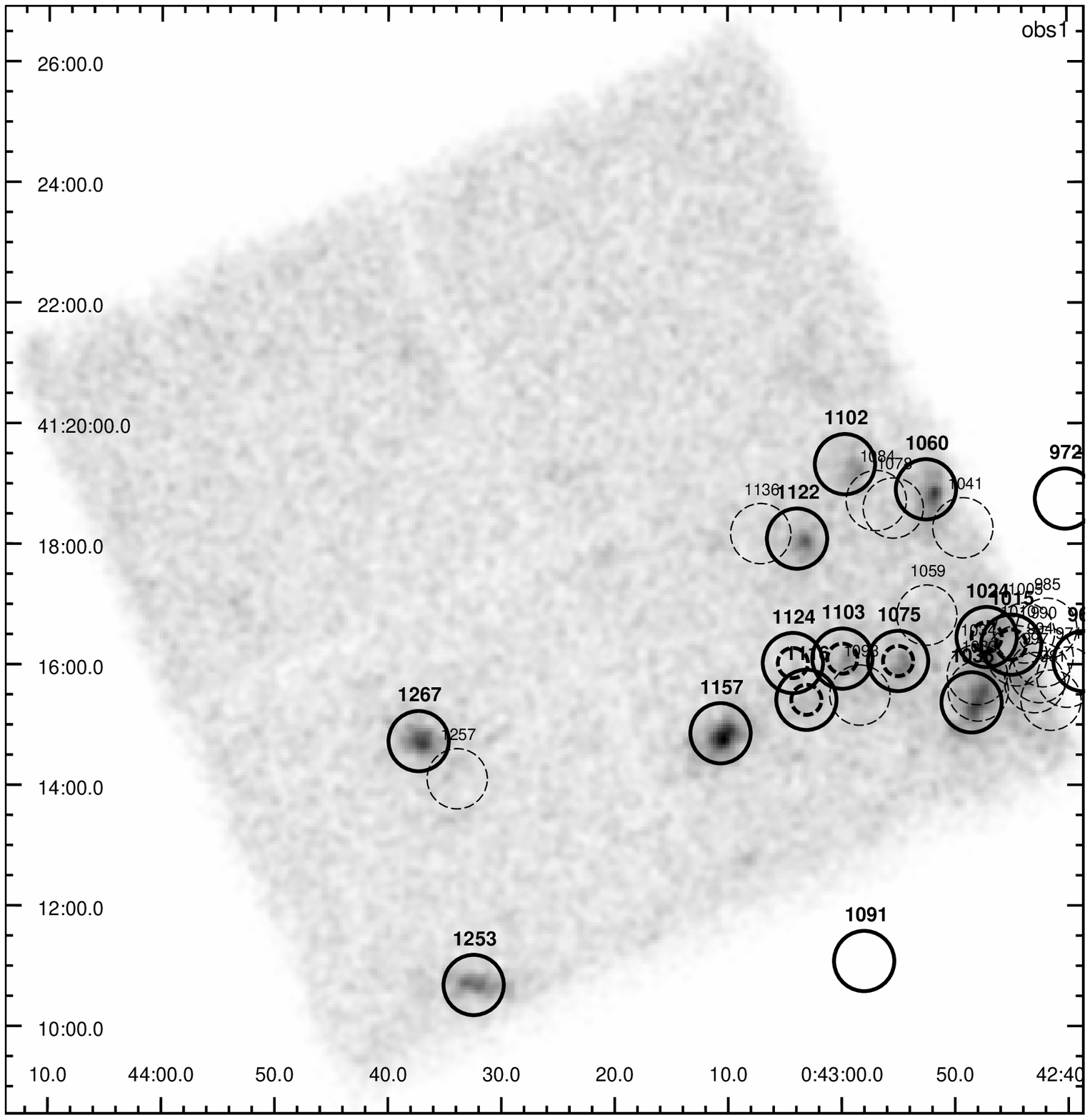}}\vspace{0.5cm}\resizebox{0.35\hsize}{!}{\includegraphics[clip,angle=0]{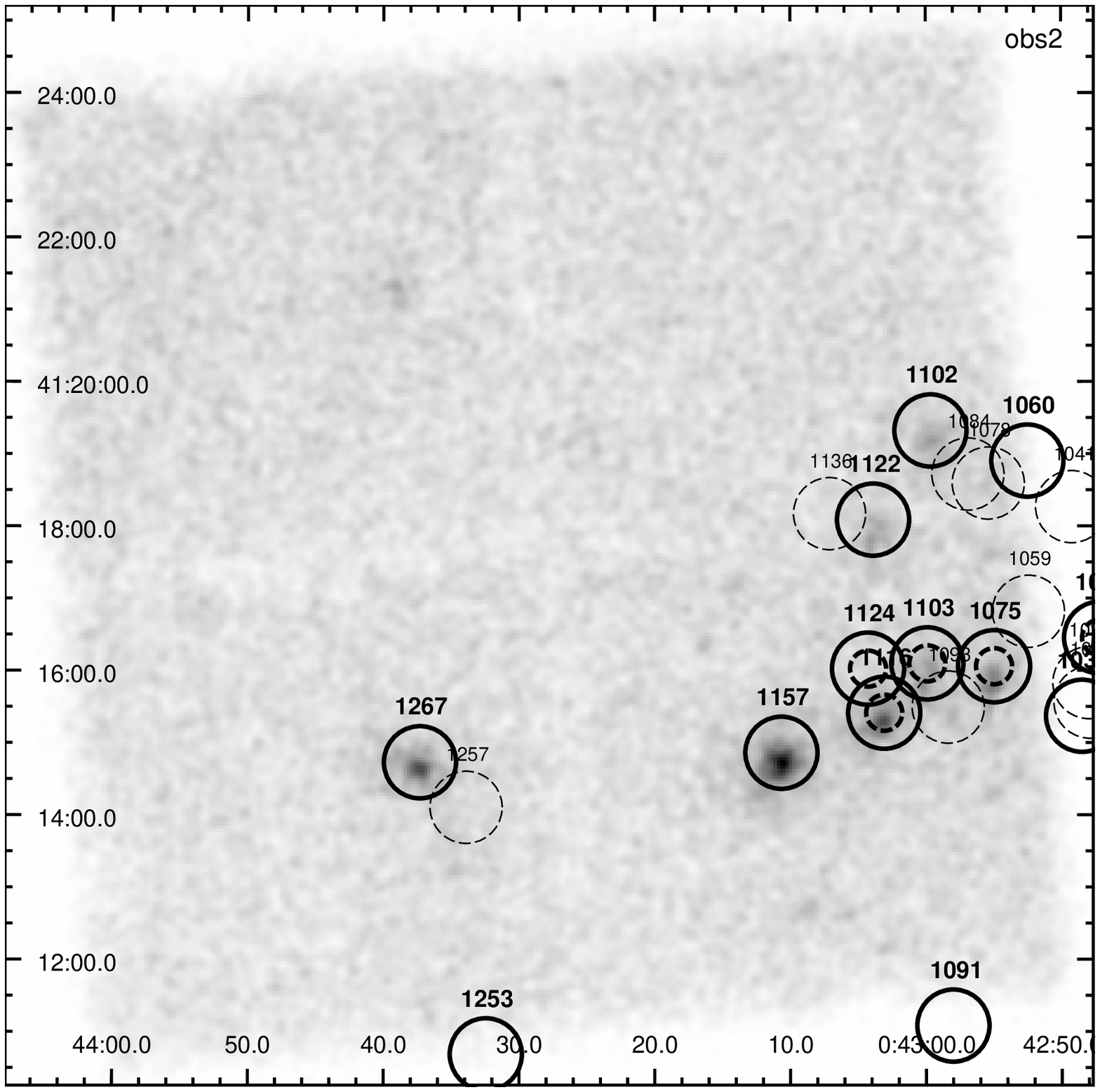}}\\
\resizebox{0.35\hsize}{!}{\includegraphics[clip,angle=0]{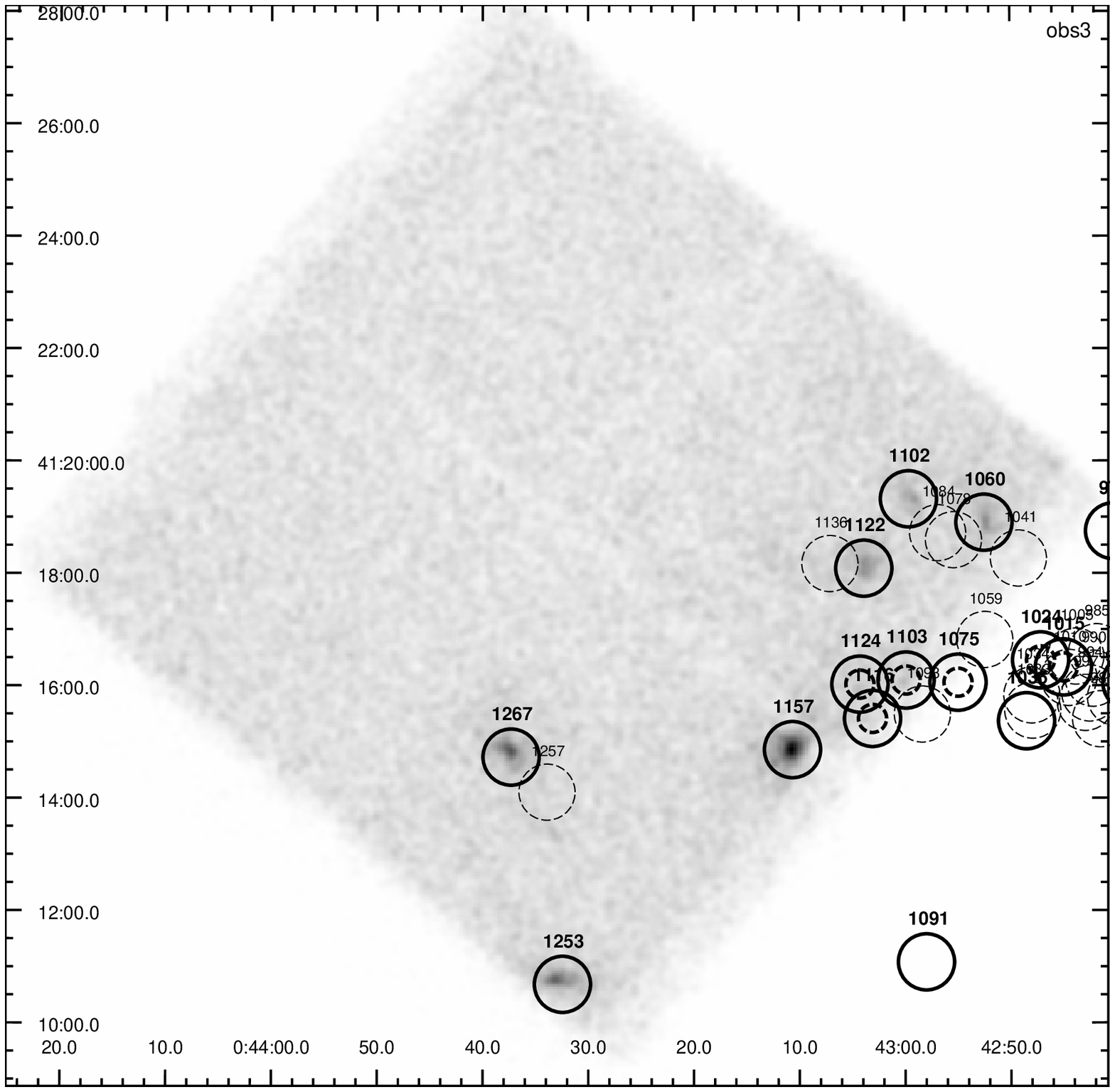}}\vspace{0.5cm}\resizebox{0.35\hsize}{!}{\includegraphics[clip,angle=0]{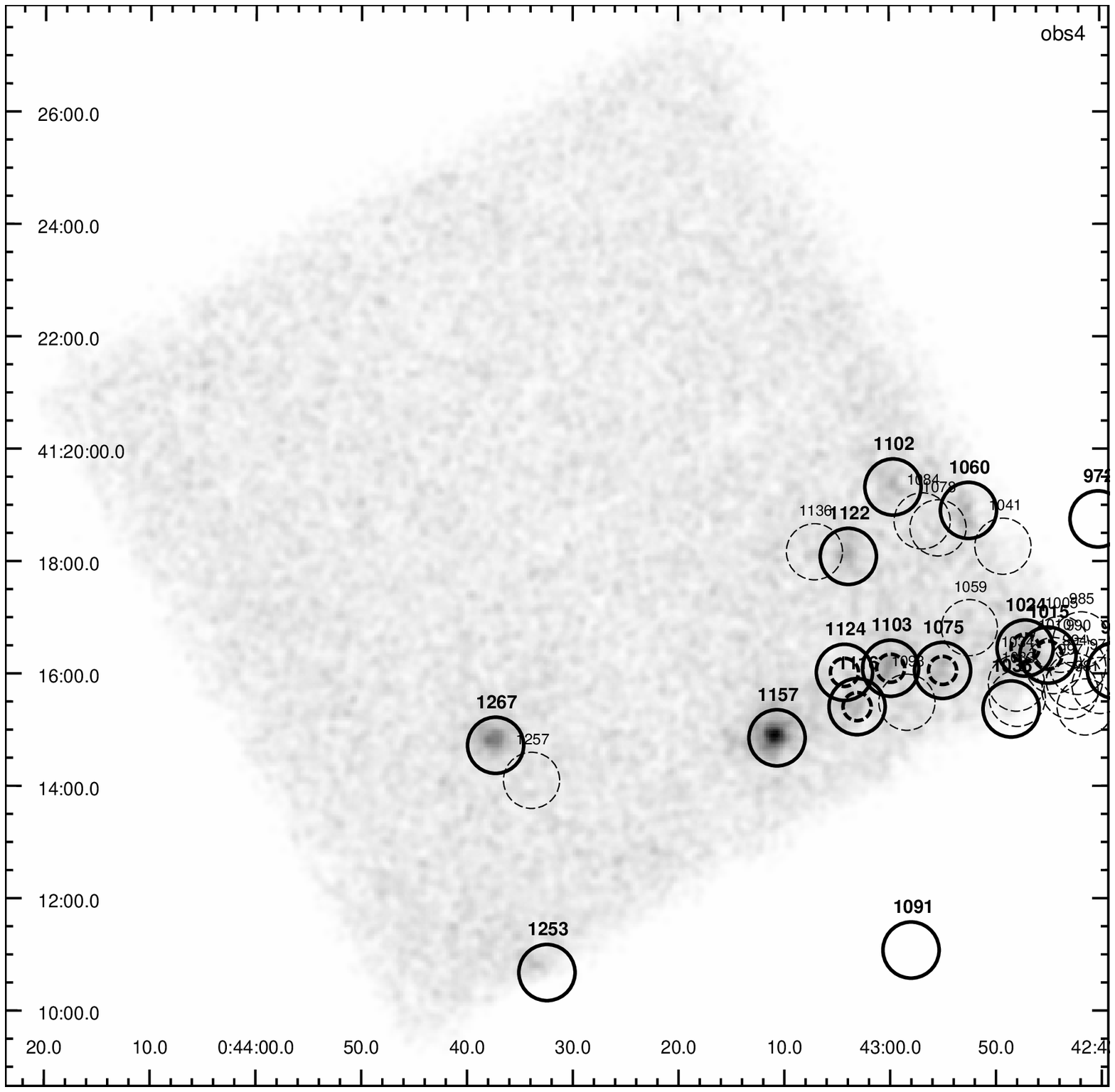}}\\
\resizebox{0.35\hsize}{!}{\includegraphics[clip,angle=0]{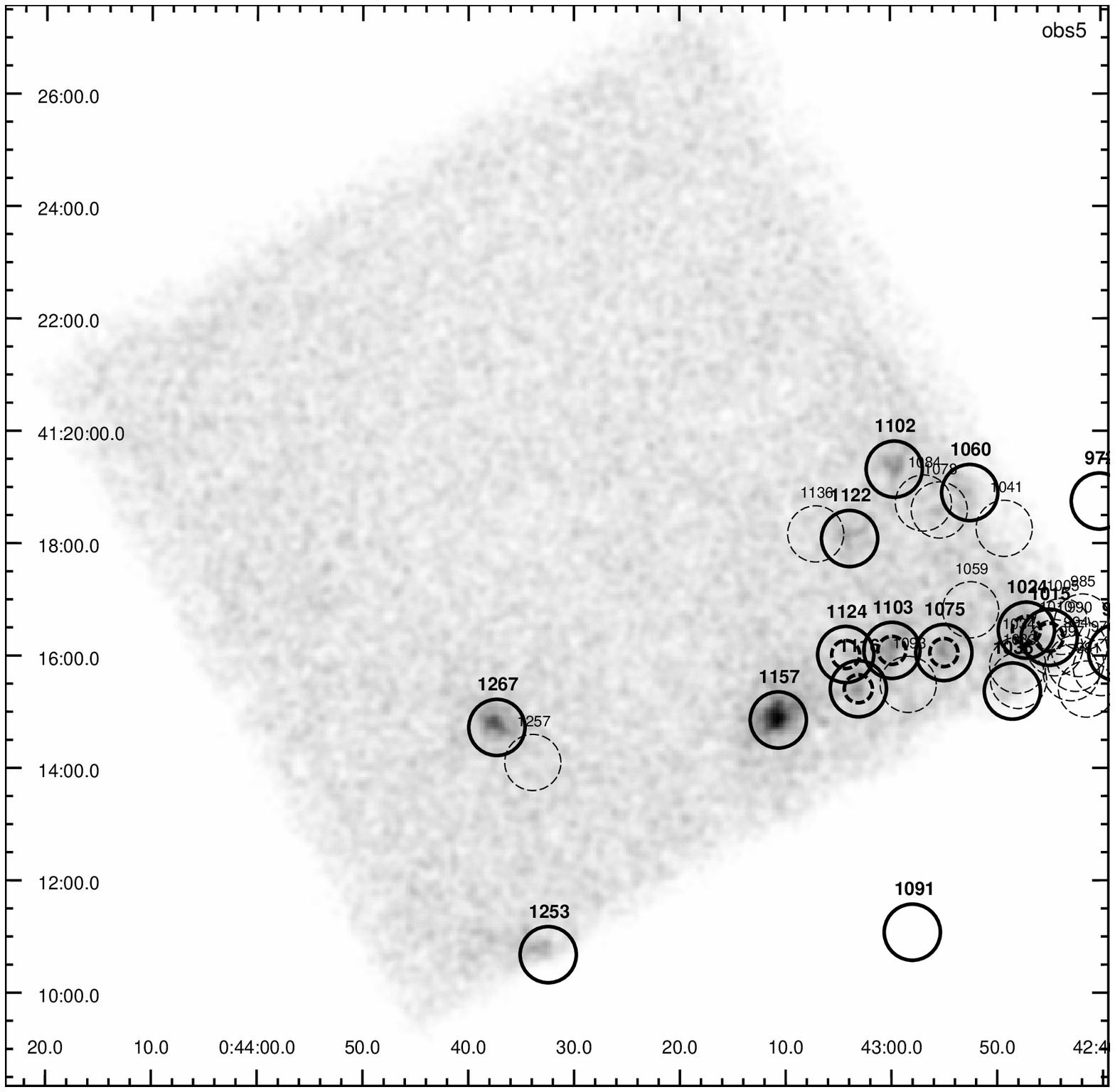}}\vspace{0.5cm}\resizebox{0.35\hsize}{!}{\includegraphics[clip,angle=0]{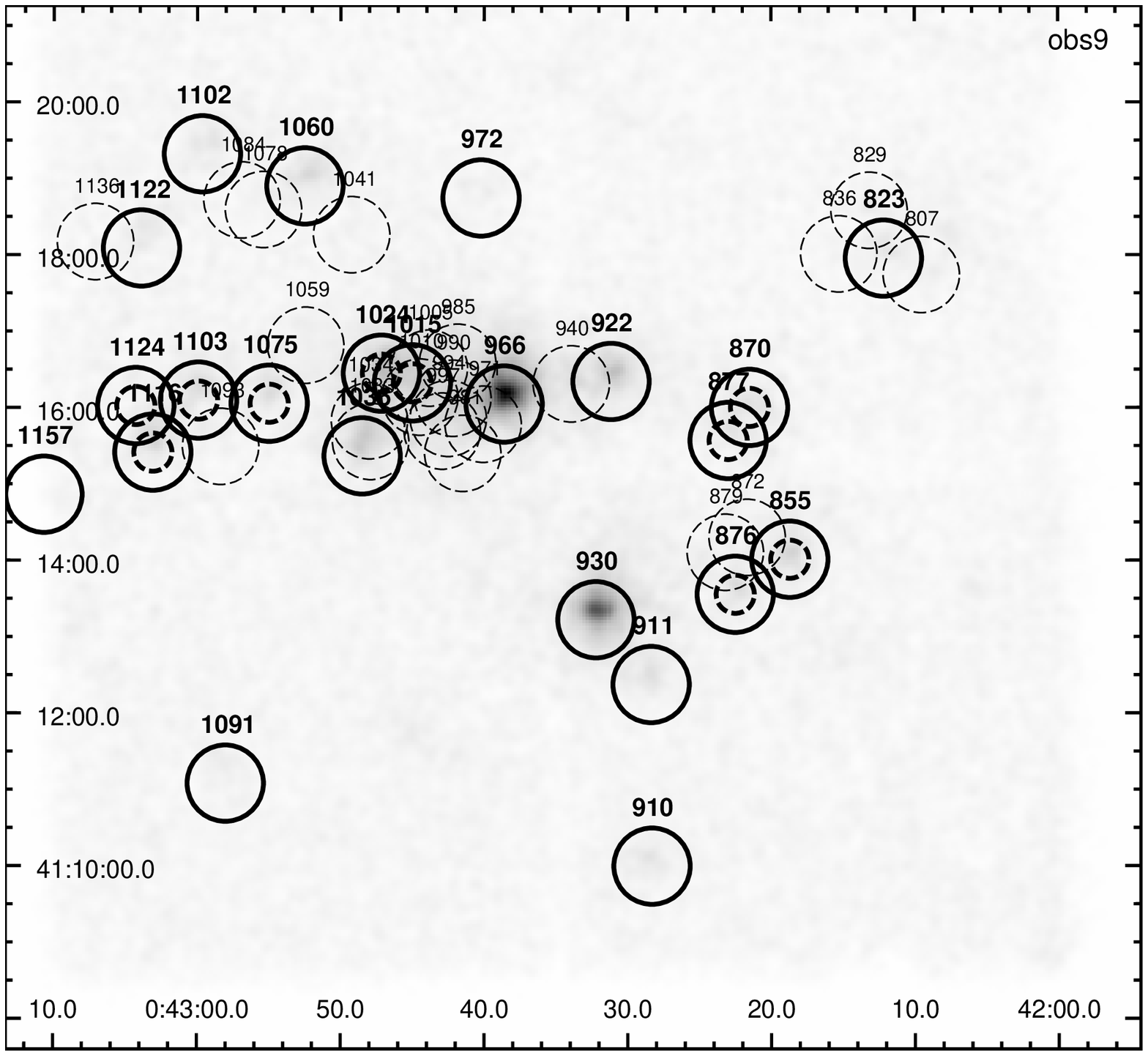}}\\
\caption{Images of the \nus/FPMB data. See Fig.~\ref{Fig:ima_nusa}.}
\label{Fig:ima_nusb}
\end{figure*}
   
\section[]{Observation and data analysis}
\label{Sec:obs}
Details on the observations used in this study are given in table \ref{Tab:data}.
We analysed the \nus\ data using the NuSTARDAS tools \texttt{nupipeline} and \texttt{nuproducts}. We loaded the cleaned images produced by \texttt{nupipeline} into \textsc{ds9} and used the source catalogue of the deep \xmm\ survey of \m31\ \citep{2011A&A...534A..55S} as a reference source list to select by eye possible \nus\ candidates of the \xmm\ sources. We also selected all sources from the \xmm\ source catalogue identified or classified as X-ray binary, globular cluster or ``hard" that are located so close to our selected \nus\ candidates that extraction regions with a radius of 30\arcsec overlap each other. For all these sources (listed in Table~\ref{Tab:srcs}) we extracted source counts from regions of 15\arcsec and 30\arcsec radius located at the known position of each source, and corresponding ``background'' counts from annulus regions with 20\arcsec$<r<$25\arcsec and 45\arcsec$<r<$54\arcsec, respectively. We then derived the ratio of the source to the background counts $R = C_{\mr{src}}/C_{\mr{bkg}}$, and the significance of the count rate difference:
\begin{equation}
Sig = \frac{C_{\mr{src}}-C_{\mr{bkg}}}{\sqrt{\sigma_{\mr{src}}^2+\sigma_{\mr{bkg}}^2}}
\end{equation}
where $\sigma_{\mr{src}}$ and $\sigma_{\mr{bkg}}$ are the errors of the source and background count rates, respectively. We derived these values for individual FPMs, as well as averages for all FPMs that cover a source position (Table~\ref{Tab:stats}), where we excluded FPMs where the source is located so close to the edge of the field of view that any of the extraction regions is only partially covered.

Images of the \nus/FPMA and B data, overlaid with source regions are shown in Figs.~\ref{Fig:ima_nusa}, \ref{Fig:ima_nusb}. Sources [SPH11] 910, 911, 930,  972, 1091, 1157, and 1253 do not overlap with any other \xmm\ source. Ignoring  all ``sources'' where $R<1$, sources [SPH11] 922, 1036, 1102, and 1267 do also not overlap with an other source. Regarding the remaining sources we included those sources in our study which have averaged significances bigger than three (including errors) for the small and big extraction regions.

For all selected sources (indicated by an \emph{italic} source ID in the first column of Table~\ref{Tab:srcs}) we extracted source photons from a circular region with a radius of 30\arcsec located at the known position of each source to derive X-ray spectra. For sources [SPH11] 855, 870, 876, 877, 1015, 1024, 1075, 1103, 1116, 1124 which overlap with at least on other source we also used the smaller radius of 15\arcsec. A circular background region with a radius of 30\arcsec located close to the source on the same detector and free of source photons was also extracted. Background subtracted source counts in the 3 -- 78 keV band obtained with \texttt{nuproducts} are given in Table \ref{Tab:nuscts}.

\section[]{Results}
\label{Sec:res}
Selecting the sources, we noticed that there are only perviously known sources, with a brightness comparable (or even brighter) than the selected sources, in the \nus\ data.  
The source IDs used in this paper are from the \xmm\ catalogue of \citet{2011A&A...534A..55S}. 

\subsection{Hardness ratio diagram}
\begin{figure}
\resizebox{\hsize}{!}{\includegraphics[clip,angle=0]{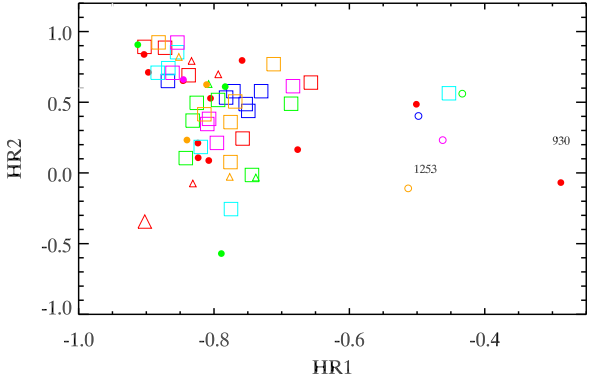}}
\caption{Hardness ratio diagram for all sources. To keep the plot less busy no error bars are shown. Error bars are given in the HR diagrams of individual sources (Fig.\ \ref{Fig:HRs1}). Counterparts of globular clusters are marked with squares, those of X-ray binaries with triangles. Big symbols indicate identified sources, while small symbols indicate candidates (as given in Table~\ref{Tab:srcs}). Hardness ratios of [SPH11] 1253 are marked with open circles. Different colours indicate different observations: obs.\ 1: magenta; obs.\ 2: cyan; obs.\ 3: blue; obs.\ 4: green; obs.\ 5: orange; obs.\ 6: red.}
\label{Fig:HRD}
\end{figure}

We used \textsc{isis} \citep[V.~1.6.2;][]{2000ASPC..216..591H} to obtain estimates of the background subtracted source count rates from the \nus\ energy spectra in band 1 (3 -- 10 keV), band 2 (10 -- 20 keV), and band 3 (20 -- 78 keV) combining data of both focal plane instruments. We derived two hardness ratios (HR) and errors for each source in each observation using the following equations:
\begin{equation}
\mr{HR}i = \frac{B_{i+1}-B_i}{B_{i+1}+B_i} \hspace{2ex}\mr{and} \hspace{2ex}
\mr{EHR}i = 2\frac{\sqrt{\lb(B_{i+1}EB_i\rb)^2+\lb(B_iEB_{i+1}\rb)^2}}{\lb(B_{i+1}+B_i\rb)^2}
\end{equation} 
for i = 1, 2, where $B_i$ and $EB_i$ denote count rates and corresponding errors in the energy band $i$.

In the HR diagram (Fig.~\ref{Fig:HRD}), sources that are already classified as X-ray binaries or globular cluster sources are located at HR1$\le-0.6$. For the 13 sources that were in the field of view of more than one observation, we show HR diagrams of each source in Fig.\ \ref{Fig:HRs1}. From these diagrams we see that for all sources the hardness ratios obtained from different observations do agree within error bars. The four observations of [SPH11] 1253 give slightly higher HR1 values and the HR1 value of [SPH11] 930 is even higher. The location of [SPH11] 930 in the HR diagram in comparison with the location of sources classified as X-ray binaries suggests that the X-ray spectral properties of [SPH11] 930 differ from those classified as X-ray binary. 
Source [SPH11] 1122 shows the strongest evolution in HR1. In obs.\ 2, it shows the highest HR1 value, which is comparable to those of [SPH11] 1253. The results of a more detailed spectral study (see Sect.\ \ref{Sec:spec}) confirm that [SPH11] 1122 is harder in obs.\ 2 than in any other observation.  Another source with HR1 values close to those of [SPH11] 1253 is [SPH11] 972, which is also fainter than [SPH11] 1253. 
In the region where the classified X-ray binaries are located there are nine sources ([SPH11] 870, 876, 877, 911, 922, 1015, 1036, 1075, 1091) that have only be classified as ``hard" in \citet{2011A&A...534A..55S}. Source [SPH11] 876 has the highest and [SPH11] 1075 the lowest HR1 value of these nine sources. Source [SPH11] 1015 has been covered by three observations, and sources [SPH11] 1036 and 1075 have been covered by four observations. The nine ``hard" sources located at HR1$\le-0.6$ are possible X-ray binary candidates.

\begin{figure*}
\resizebox{\hsize}{!}{\includegraphics[clip,angle=0]{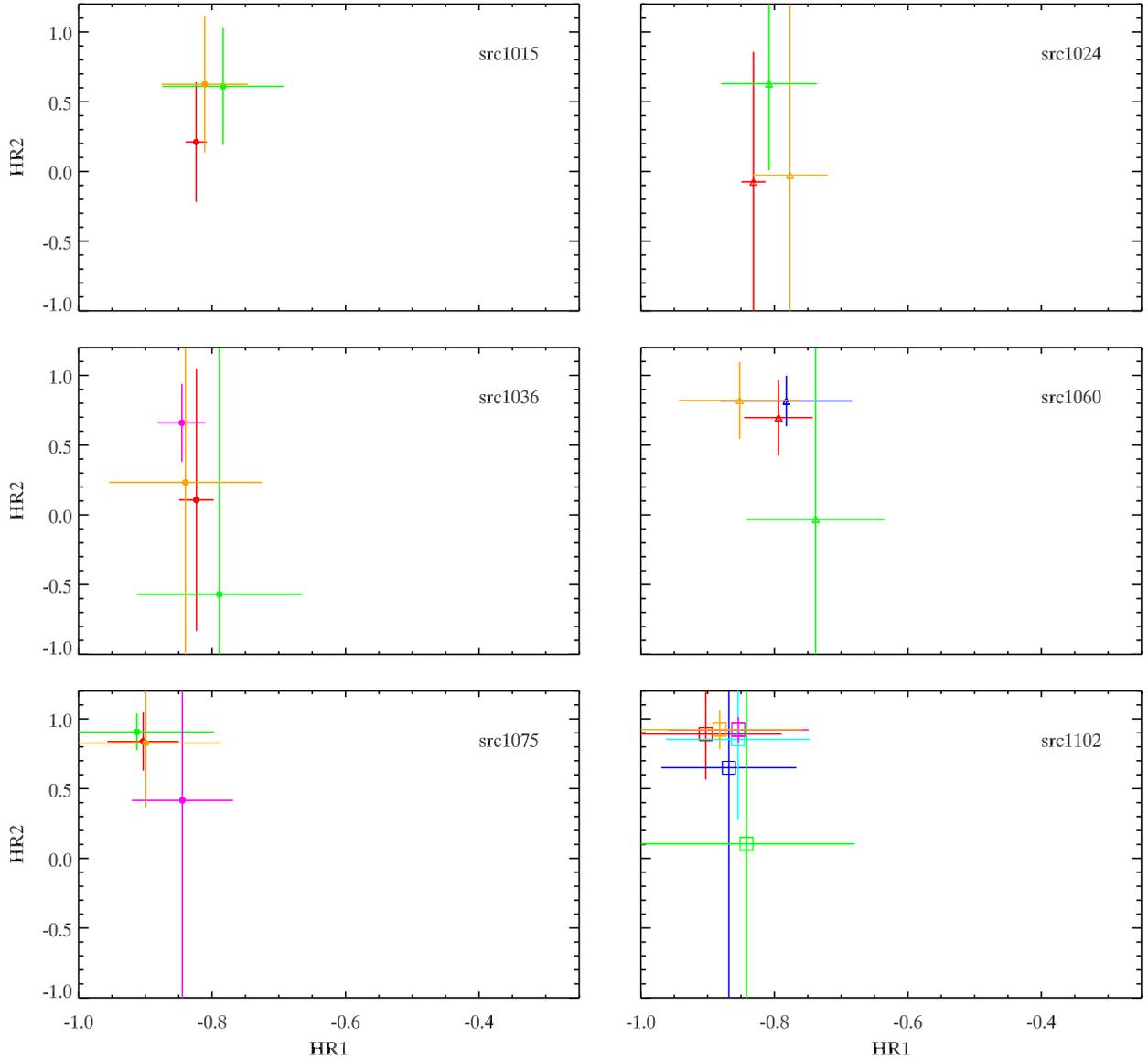}}
\caption{Hardness ratio diagrams of the 13 sources that were observed more than once. Different colours indicate different observations; different symbols indicate different source types (see Fig.\ \ref{Fig:HRD}).}
\label{Fig:HRs1}
\end{figure*}

\addtocounter{figure}{-1}
\begin{figure*}
\resizebox{\hsize}{!}{\includegraphics[clip,angle=0]{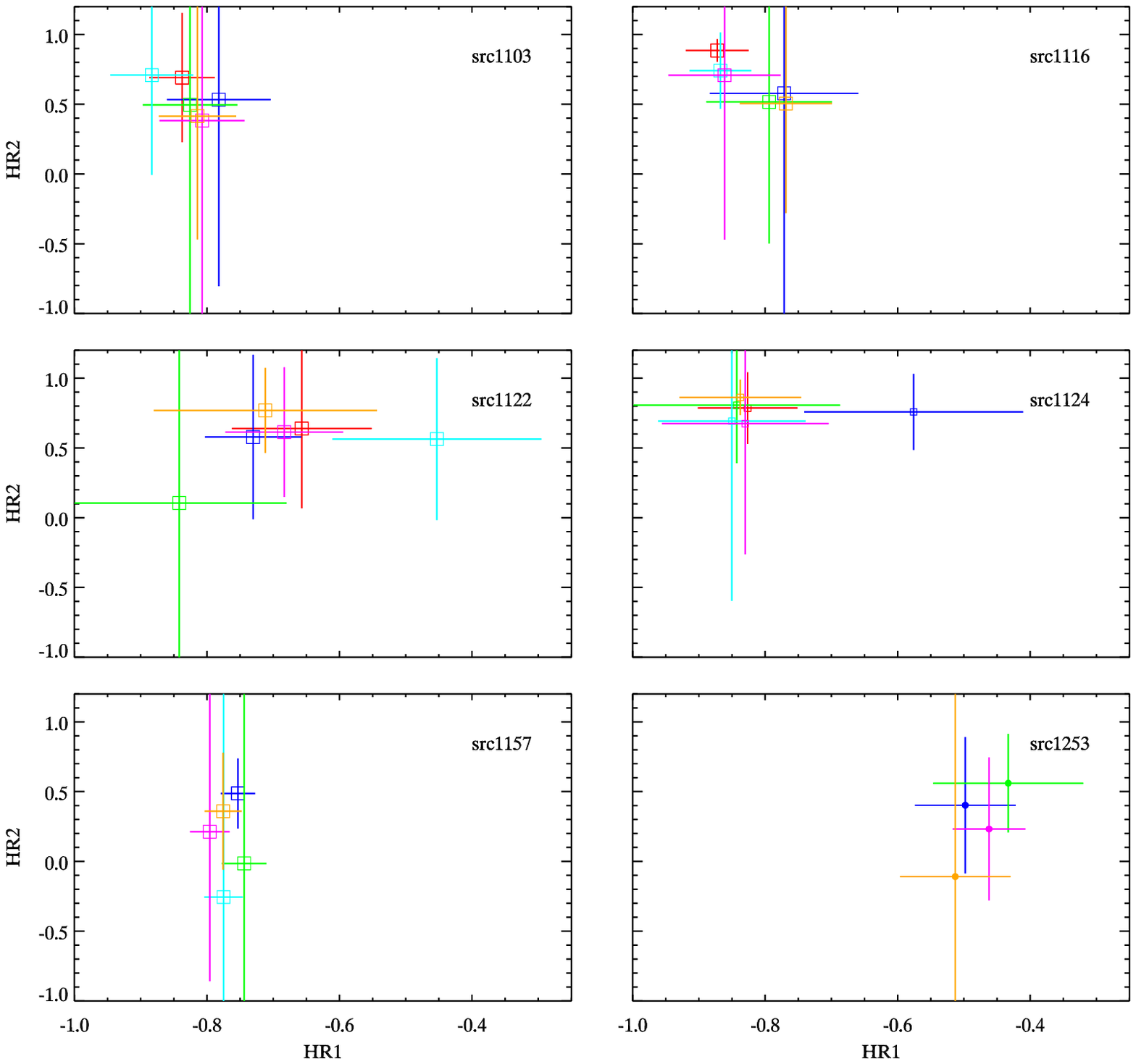}}
\caption{continued}
\label{Fig:HRs2}
\end{figure*}

\addtocounter{figure}{-1}
\begin{figure}
\resizebox{\hsize}{!}{\includegraphics[clip,angle=0]{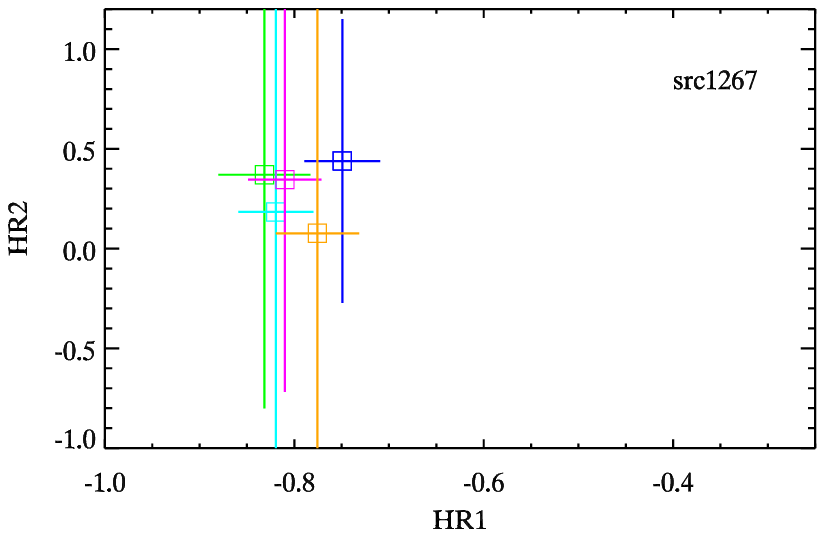}}
\caption{continued}
\label{Fig:HRs3}
\end{figure}

\begin{table*}
\caption{\label{Tab:Spec_par}Spectral parameters obtained by fitting the \nus\ spectra. Luminosities in units of $10^{36}$ erg/s. This is a sample of the full table, which is available with the electronic version of the article.}
\begin{center}
\tiny
\begin{tabular}{rrllllllllll}
\hline\noalign{\smallskip}
\multicolumn{1}{c}{src.} &  \multicolumn{1}{c}{obs.} &  \multicolumn{1}{c}{$\Gamma$} & \multicolumn{1}{c}{N$_{\mr{pl}}$} & \multicolumn{1}{c}{C$_{\mr{FPMB; pl}}$} & \multicolumn{1}{c}{T$_{\mr{in}}$ [keV]} &  \multicolumn{1}{c}{N$_{\mr{dbb}}$} & \multicolumn{1}{c}{C$_{\mr{FPMB; dbb}}$} & \multicolumn{1}{c}{T$_{\mr{0}}$ [keV]} & \multicolumn{1}{c}{T$_{\mr{e}}$ [keV]} & \multicolumn{1}{c}{$\tau$}&  \multicolumn{1}{c}{N$_{\mr{ctt}}$} \\
& & \multicolumn{1}{c}{C$_{\mr{FPMB; ctt}}$} &  \multicolumn{1}{c}{$\chi^2_{\mr{pl}}/dof$}  &  \multicolumn{1}{c}{$\chi^2_{\mr{dbb}}/dof$}  &  \multicolumn{1}{c}{$\chi^2_{\mr{ctt}}/dof$}  &  \multicolumn{1}{c}{$L_{\mr{pl}}$} &  \multicolumn{1}{c}{$L_{\mr{dbb}}$} &  \multicolumn{1}{c}{$L_{\mr{ctt}}$} & & &\\
\hline\noalign{\smallskip}
 823&6 & $2.40_{-0.24}^{+0.25} $ &$3.51_{-1.29}^{+2.00}E-04 $ &$0.86_{-0.15}^{+0.19} $ &$2.40_{-0.33}^{+0.41} $ &$8.70_{-4.24}^{+7.95}E-04 $ &$0.87_{-0.16}^{+0.19} $ &$4.76_{-4.76}^{+4.53}E-01 $ &$2.89_{-0.97}^{+84.01} $ &$ 5.52_{-5.52}^{+4.48} $ &$6.12_{-6.12}^{+329.60}E-05 $ \\
 & &$0.86_{-0.16}^{+0.19} $ &60.15/58& 61.02 /58& 58.24/56& $3.78_{-0.37}^{+0.37} $ &$	2.37_{-0.23}^{+0.23} $ &$	2.61_{-0.26}^{+0.26} $ & & & \\ 
 855& 6& $ 2.39_{-0.18}^{+0.19} $ &$ 3.86_{-1.16}^{+1.64}E-04 $ &$ 0.94_{-0.15}^{+0.18}$ & $ 2.80_{-0.34}^{+0.40}$ & $ 5.42_{-2.30}^{+3.91}E-04$ & $ 0.91_{-0.15}^{+0.17}$ & $ 2.02_{-2.02}^{+6.47}E-03$ & $ 3.31_{-0.35}^{+0.39}$ & $ 5.30_{-0.56}^{+0.66} $ & $ 8.00_{-0.71}^{+538020.00}E-03$\\
&  & $ 0.92_{-0.15}^{+0.17}$ & 31.79/ 33& 28.27/ 33& 24.73/ 31& $4.33_{-0.37}^{+0.37} $ &$	2.96_{-0.25}^{+0.25} $ &$	3.22_{-0.28}^{+0.28} $ & & & \\
 870&6 & $2.45_{-0.14}^{+0.15} $ &$4.27_{-1.00}^{+1.29}E-04 $ &$1.10_{-0.13}^{+0.15} $ &$2.40_{-0.20}^{+0.22} $ &$1.06_{-0.34}^{+0.48}E-03 $ &$1.02_{-0.12}^{+0.14} $ &$1.43_{-1.43}^{+10.04}E-01 $ &$2.14_{-0.38}^{+0.67} $ &$8.12_{-2.44}^{+4.43} $ &$2.00_{-0.19}^{+1668.80}E-03 $\\
&  &$1.03_{-0.13}^{+0.15} $ &105.80/79 &87.77/79& 87.47/77& $4.16_{-0.27}^{+0.27} $ &$	2.89_{-0.18}^{+0.18} $ &$	2.91_{-0.19}^{+0.19} $ & & & \\ 
 876& 6& $ 2.06_{-0.26}^{+0.28} $ &$ 1.28_{-0.56}^{+0.99}E-04 $ &$ 0.95_{-0.20}^{+0.26}$ & $ 3.79 _{-0.76}^{+1.02}$ & $ 9.33_{-5.68}^{+13.90}E-05$ & $ 0.98_{-0.21}^{+0.27}$ & $ 4.64_{-4.64}^{+5.19}E-01$ & $ 5.99_{-2.75}^{+160.50 }$ & $ 3.86_{-3.80}^{+3.00} $ & $ 1.71_{-1.71}^{+59.76}E-05$\\
&  & $ 0.95_{-0.20}^{+0.26}$ & 21.81/ 24& 26.74/ 24& 21.17/ 22& $3.26_{-0.37}^{+0.37} $ &$	1.93_{-0.22}^{+0.22} $ &$	2.42_{-0.28}^{+0.28} $ & & & \\
 877&6& $2.35_{-0.21}^{+0.23} $ &$2.80_{-0.94}^{+1.39}E-04 $ &$1.03_{-0.18}^{+0.22} $ &$2.69_{-0.36}^{+0.44} $ &$4.83_{-2.22}^{+3.98}E-04 $ &$1.01_{-0.18}^{+0.21} $ &$9.95_{-9.95}^{+1006.89}E-03 $ &$3.26_{-1.03}^{+7.42} $ &$5.26_{-3.09}^{+3.27} $ &$1.66_{-1.66}^{+0.30}E-03 $ \\
& &$1.02_{-0.18}^{+0.21} $ &82.51/74& 81.92/74 &78.62/72 &$3.40_{-0.32}^{+0.32} $ &$	2.23_{-0.21}^{+0.21} $ &$	2.43_{-0.23}^{+0.23} $  & & & \\ 
\hline\noalign{\smallskip}
\end{tabular}
\end{center}
\end{table*}

\subsection{Spectral properties}
\label{Sec:spec}
\begin{figure*}
\resizebox{0.7\hsize}{!}{\includegraphics[clip,angle=0]{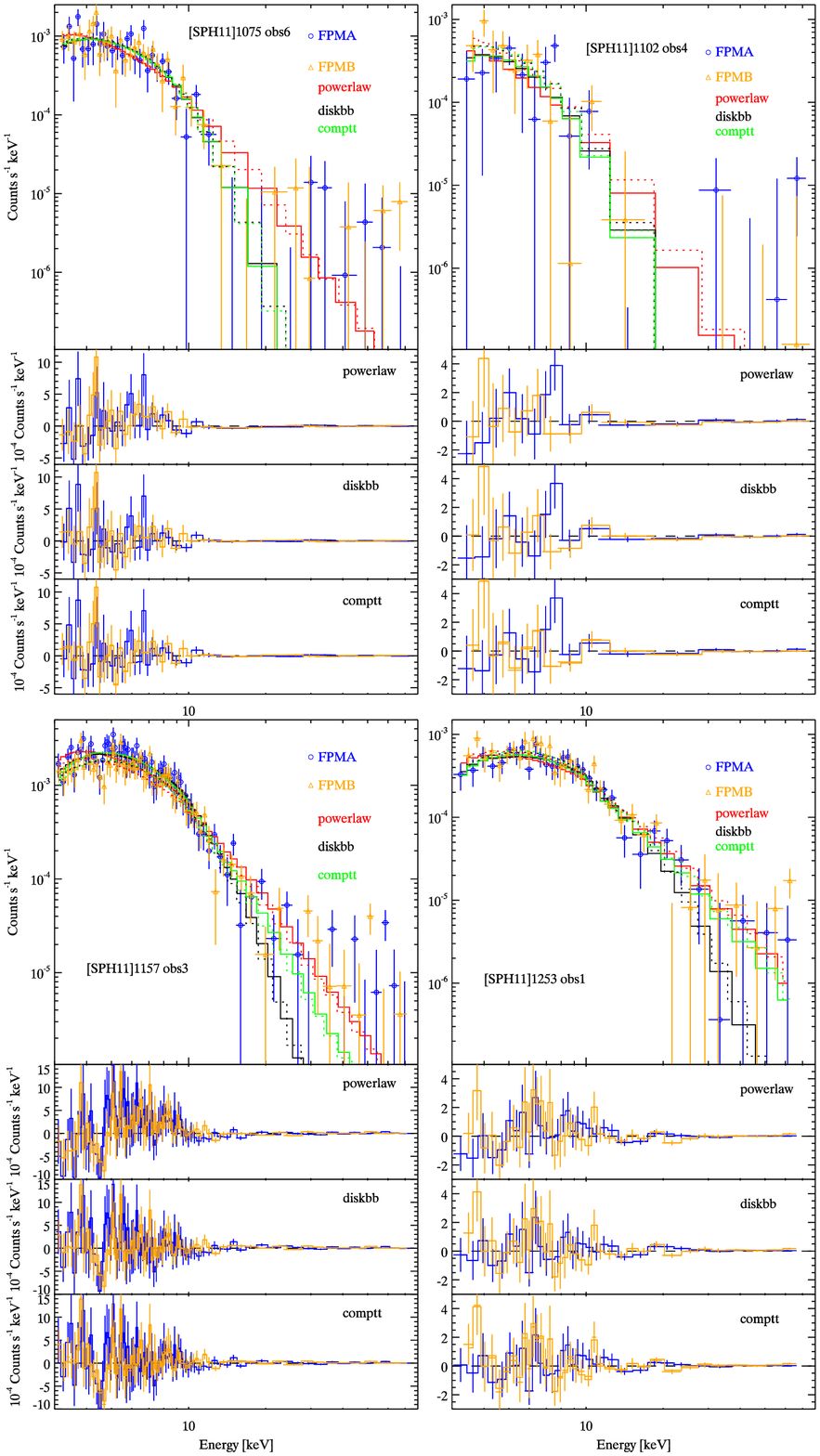}}
\caption{Spectra of [SPH11]1075, 1102, 1157, and 1253. For each source the spectra with the best fit of each model and the residuals for each model are shown.}
\label{Fig:spec}
\end{figure*}

To investigate this possibility further, we fitted simultaneously the FPMA and FPMB energy spectra of each source within \textsc{Xspec} \citep[V.~12.8.2;][]{1996ASPC..101...17A} in the 3 -- 78 keV range, grouping the  data to a minimum of 20 counts in each bin. We use the following three simple models, which have also been used in \citet{2016MNRAS.458.3633M}, to fit the data: \texttt{powerlaw}, \texttt{diskbb} \citep{1984PASJ...36..741M}, and \texttt{comptt}  \citep{1994ApJ...434..570T}, including Galactic foreground absorption using \texttt{TBabs} \citep{2000ApJ...542..914W}. The obtained best fit parameters are given in Table~\ref{Tab:Spec_par}, together with luminosities and errors, derived using the \texttt{cflux} model. Four spectra with the best fit models and residuals are shown in Fig.~\ref{Fig:spec}. We selected the brightest observations of [SPH11] 1075, 1157 (the brightest globular cluster source in our sample) and 1253, and the faintest observation of 1102 (the faintest globular cluster source in our sample). We find that most sources have a photon index above or close to two, similar to what has been found in \citet{2016MNRAS.458.3633M}. Exceptions are [SPH11] 930 with $\Gamma=1.22\pm0.04$ and [SPH11] 1253 with a photon index around 1.5.\@ We can also confirm another result of \citet{2016MNRAS.458.3633M}, that for most observations the \texttt{comptt} model provides a statistically better fit than a power law. 

\begin{figure}
\resizebox{\hsize}{!}{\includegraphics[clip,angle=0]{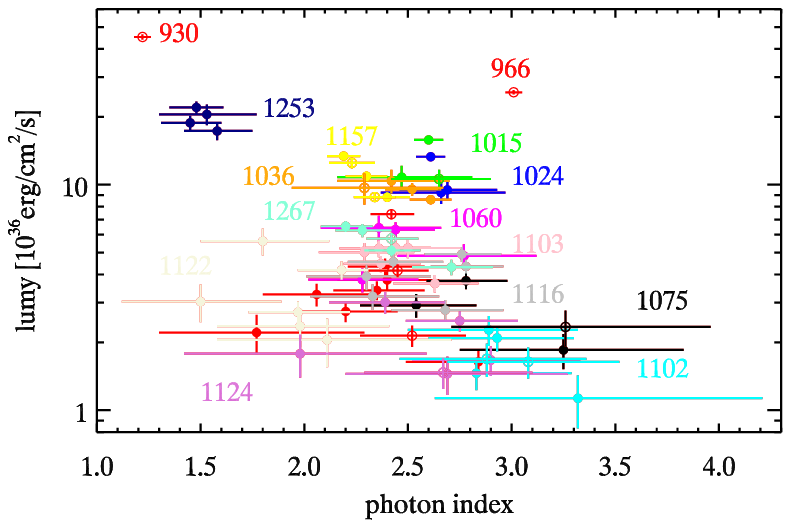}}
\caption{Correlation between photon index and source luminosity.}
\label{Fig:phI_lumy}
\end{figure}

For sources observed more than once we can investigate the evolution of the spectral parameters. For almost all observations the value of the cross-calibration constant is independent of the spectral model used. For most sources the normalisations of the power law and of the disk blackbody show a similar evolution, and the evolution of the photon index and inner disc temperature is anti-correlated. The photon index and inner disc temperature and radius are constant within errors. Exceptions are  [SPH11]1122 (obs2/3) and [SPH11]1267 (obs4). In general, the seed temperature in the \texttt{comptt} model is lower than the disc temperature in the \texttt{diskbb} model and it shows a different evolution. While a disc temperature of $\sim2$ keV is quite high compared to values obtained from Galactic X-ray binaries using multi-component spectral models, it is in agreement with the values reported in \citet{2016MNRAS.458.3633M}. Regarding correlations between spectral parameters and source flux, we do not find any overall correlation of any specific spectral parameter with flux. The evolution of the photon index with luminosity is shown in Fig.~\ref{Fig:phI_lumy}. Only [SPH11]1157 and 1267 are getting harder when they get brighter. For these two sources we also find that the disc temperature increases when they get brighter. [SPH11]1267 shows a decreasing radius with increasing flux, while the radius of [SPH11]1157 is constant. 

\subsection{Long term light curves}
For the 13 sources that were in the field of view of more than one observation, we prepared long term light curves. We derived the fluxes from the spectral fits adding the \texttt{cflux} component to each spectral model with all parameters fixed to their best fit value. Fluxes are derived in the 3 -- 78 keV band and converted to luminosities assuming a distance of 780 kpc. The obtained light curves are show in Fig.\ \ref{Fig:lc1}. Following \citet{2011A&A...534A..55S} we determine the variability factor, the ratio of the maximum and minimum luminosity, $Var = L_{\mr{max}}/L_{\mr{min}}$, and the significance of the luminosity difference:
\begin{equation}
Sig = \frac{L_{\mr{max}}-L_{\mr{min}}}{\sqrt{\sigma_{\mr{max}}^2+\sigma_{\mr{min}}^2}}
\end{equation}
where $\sigma_{\mr{max}}$ and $\sigma_{\mr{min}}$ are the errors of the maximum and minimum luminosity (ignoring uncertainties on the distance), respectively. 
Apart from the brightest source ([SPH11]1253) the overall evolution of the luminosity with time is consistent between the different models. For [SPH11] 1015, 1024, 1060, 1102, 1103, 1122, 1124, 1157  and 1267 we observe an overall flux increase, while [SPH11] 1036 and 1253 show an overall decreasing flux. For [SPH11] 1102 and 1122 there are two observations (obs.\ 4, 5) where the flux deviates from the overall increase, as it is lower in these two observations. For [SPH11] 1103 and 1157 the flux in the first two observations is lower than in the remaining observations. However, the variability factor is rather low. For most sources the variability factor is about 1 -- 2, and it is below 5 for all sources. For most sources the significance of variability is $\la 3$, only [SPH11] 1157 has $Sig> 5$. For [SPH11] 1015, 1024, 1075, 1103, 1116, and 1124 we also show luminosities derived from source extraction regions of 15\arcsec. For most observations the overall evolution of the luminosity does not depend on the size of the extraction region. An exception is [SPH11] 1124 where the evolution of the luminosity in the last four observations does not agree with each other. For most observations the luminosities derived from the 15\arcsec\ regions are lower than those derived from the 30\arcsec\ regions. In case the evolution of the luminosities does not depend on the size of the extraction region, the lower luminosities from the 15\arcsec\ regions imply that we miss source flux using these extraction regions, while the bigger extraction regions are still dominated by the source flux and contribution from nearby sources does not seem to affect the fluxes obtained. 

\begin{figure*}
\resizebox{\hsize}{!}{\includegraphics[clip,angle=0]{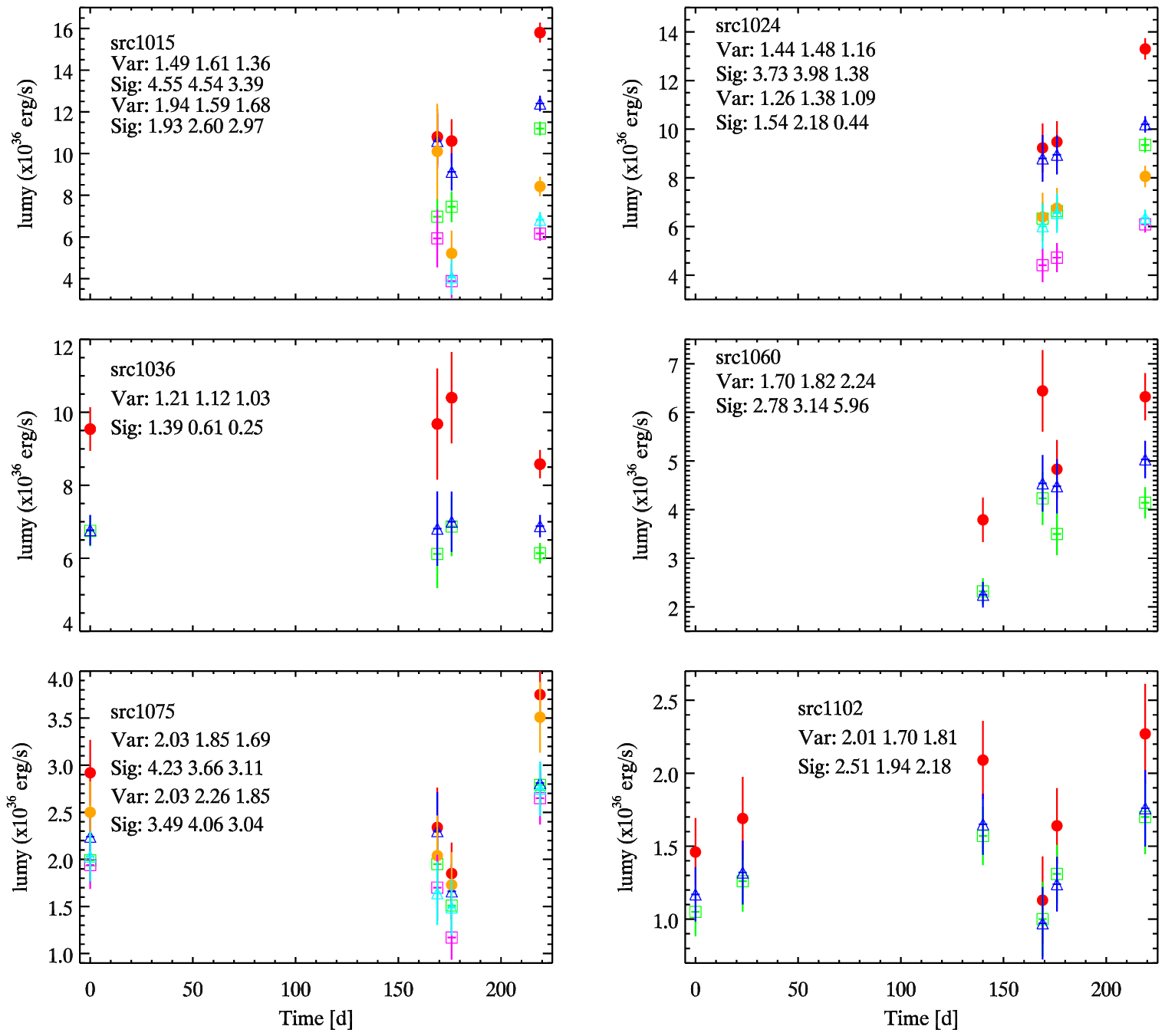}}
\caption{Long term light curves of the 13 sources that were observed more than once, derived form the fits of the energy spectra. Luminosities obtained from the power law fit are given as red filled dots, those from the disk blackbody fit as green open squares, and those form the Comptonization model as blue open triangles. The ratio of the maximum and minimum luminosity (Var) and the significance of the luminosity difference (Sig) are given for each spectral model (power law, disk blackbody, Comptonization model). For [SPH11] 1015, 1024, 1075, 1103, 1116, and 1124 we also show luminosities derived from extraction regions of 15\arcsec\ (power law: orange dots; disc blackbody: magenta squares; Comptonization model: cyan triangles).}
\label{Fig:lc1}
\end{figure*}

\addtocounter{figure}{-1}
\begin{figure*}
\resizebox{\hsize}{!}{\includegraphics[clip,angle=0]{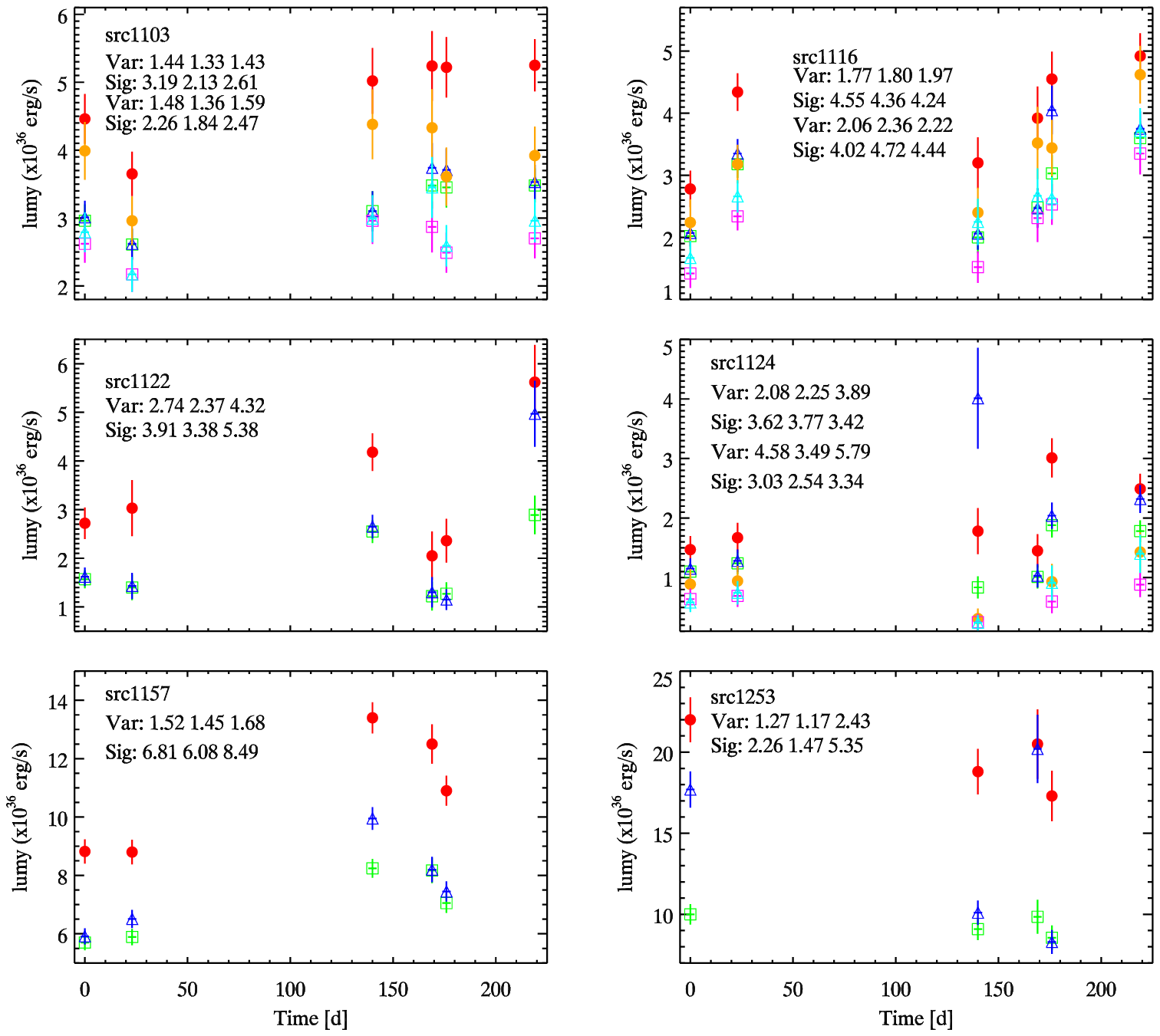}}
\caption{continued}
\label{Fig:lc2}
\end{figure*}

\addtocounter{figure}{-1}
\begin{figure}
\resizebox{\hsize}{!}{\includegraphics[clip,angle=0]{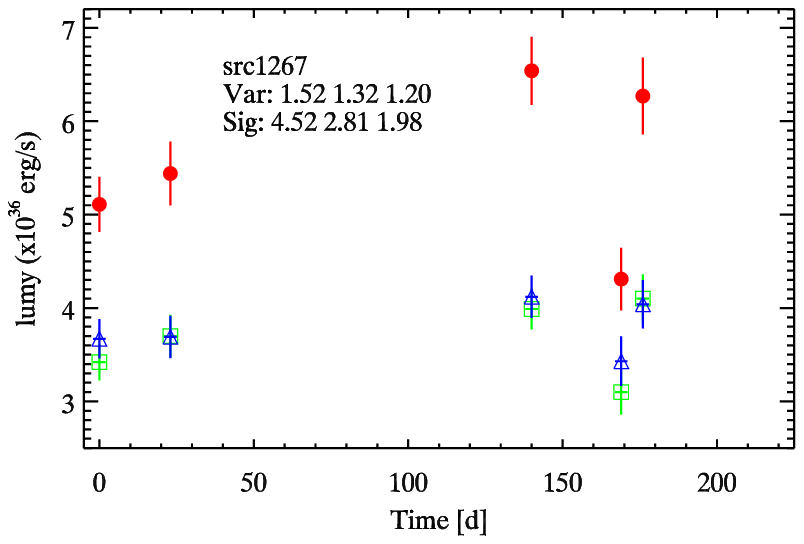}}
\caption{continued}
\label{Fig:lc3}
\end{figure}

\section[]{Discussion}
\label{Sec:dis}

We analysed \nus\ data of \m31\ taken between February and September 2015. Comparing with the \xmm\ source list presented in \citet{2011A&A...534A..55S} we find that all sources visible in the \nus\ data have already been seen with \xmm, \ie\ the \nus\ data do not contain any transient source that was not active at the time of the \xmm\ data analysed by \citet{2011A&A...534A..55S}. Investigating hardness ratios we find that most sources are located in a certain region of the hardness ratio diagram. More detailed spectral fitting confirms that most sources have similar spectral properties. This suggest that [SPH11] 922, 1015, 1036, 1075, which have only been classified as  ``hard" in \citet{2011A&A...534A..55S} are good candidates for being X-ray binaries, as their spectral properties are quite similar to the ones of the identified X-ray binaries included in our source sample. 

\begin{figure}
\resizebox{\hsize}{!}{\includegraphics[clip,angle=0]{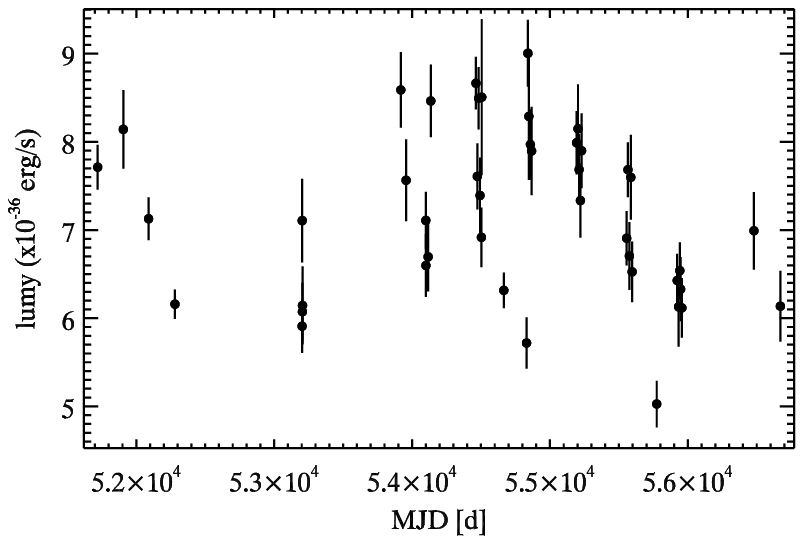}}
\caption{Long term \xmm\ light curve of [SPH11] 1253.}
\label{Fig:lc_xmm}
\end{figure}

Sources [SPH11] 930 and [SPH11] 1253 are located at higher HR1 values than the identified X-ray binaries, and our more detailed spectral investigation confirms that these two sources have harder spectra than the identified X-ray binaries. There are several possibilities why the spectra of sources [SPH11] 930 and [SPH11] 1253 differ from those of the identified X-ray binaries. One possibility is that these two sources are background AGN, not belonging to the X-ray binary population of \m31. Another possibility is that these two sources are X-ray binaries observed in a (much) harder state than the identified X-ray binaries. \citet{2013A&A...555A..65H} suggest [SPH11] 930 to be an X-ray binary candidate based on the long-term variability observed in \textit{Chandra} HRC-I observations. \citet{2017ApJ...838...47Y} reported the identification of a bright hard X-ray source dominating the \m31\ bulge above 25 keV (the \nus\ counterpart of [SPH11] 930), which is the counterpart to Swift J0042.6+4112, a possible X-ray pulsar with an intermediate-mass companion or a symbiotic X-ray binary. \citet{2017ApJ...851L..27M} reanalysed archival \xmm\ observation of this source and discovered periodic dips. They suggest [SPH11] 930 to be a dipping low-mass X-ray binary seen at high inclination. 
\citet{2016MNRAS.458.3633M} argue that X-ray binaries associated with globular clusters in \m31\ are most likely containing a neutron star primary. So the harder spectra of [SPH11] 1253 can indicate that this source is a black hole X-ray binary. A photon index of $\sim$1.5, as seen in [SPH11] 1253, can be observed in the low-hard state of a black hole X-ray binary \citep[\eg\ ][]{2013MNRAS.429.2655S,2016MNRAS.459.4038S}. As [SPH11] 1253 is a field source the harder spectra of this source may indicate that it is a (neutron star) high-mass X-ray binary. However, it is not among the list of candidate high-mass X-ray binaries of \citet[][Table 5]{2009A&A...495..733S}. The long-term \xmm\ light curve of [SPH11] 1253 obtained from the 0.2 -- 12 keV band flux given in the 3XMM-DR6 catalogue \citep{2016A&A...590A...1R} is shown in Fig.\ \ref{Fig:lc_xmm}. The light curve shows low (var = 1.79), but significant (sig: 8.62) variability. An interpretation as a supernova remnant is unlikely as the source shows significant variability and it is not included in the list of supernova remnants and candidates presented in \citet{2012A&A...544A.144S}. 

\section*{Acknowledgments}
This project is supported by the Ministry of Science and Technology of
the Republic of China (Taiwan) through grants 103-2628-M-007-003-MY3, 104-281-M-007-060, 105-2112-M-007-033-MY2 and 105-2811-M-007-065.
This research has made use of data obtained through the High Energy Astrophysics Science Archive Research Center Online Service, provided by the NASA/Goddard Space Flight Center. This research has made use of data obtained from the 3XMM XMM-Newton serendipitous source catalogue compiled by the 10 institutes of the XMM-Newton Survey Science Centre selected by ESA.

\bibliographystyle{mnras}
\bibliography{}


\bsp

\label{lastpage}

\end{document}